\documentclass[fleqn,usenatbib]{mnras}

\usepackage[T1]{fontenc}
\usepackage{xcolor}
\usepackage{amsmath,amssymb}
\usepackage{graphics,graphicx}
\usepackage{natbib}
\DeclareRobustCommand{\VAN}[3]{#2}
\let\VANthebibliography\thebibliography
\def\thebibliography{\DeclareRobustCommand{\VAN}[3]{##3}\VANthebibliography}

\usepackage{graphicx}	
\usepackage{amsmath}	

\newcommand{\mach}{\mathcal{M}}

\title[The origin of supermassive black holes at  cosmic dawn]{The origin of supermassive black holes at  cosmic dawn}

\author[]{
	Ritik Sharma$^{\href{https://orcid.org/0009-0001-3469-9812}{\includegraphics[width=0.4cm]{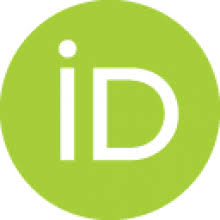}}}$ \&
Mahavir Sharma$^{\href{https://orcid.org/0000-0001-9927-5255}{\includegraphics[width=0.4cm]{orcid.jpg}}}$\thanks{E-mail: mahavir@iitbhilai.ac.in}
\\
Department of Physics, Indian Institute of Technology (IIT) Bhilai, 491001, India\\
}

\date{Accepted ..... Received ....; in original form ....}

\pubyear{2024}

\begin{document}
\label{firstpage}
\pagerange{\pageref{firstpage}--\pageref{lastpage}}
\maketitle

\begin{abstract}
We investigate the steady spherically symmetric accretion in the combined potential of a central black hole and a dark matter halo. For the halo, we consider a Hernquist and an NFW potential and calculate the critical points of the flow. We find that the trans-sonic solution to the centre is not possible without a black hole, whereas two types of trans-sonic solutions are possible in its presence. We also derive the mass accretion rate for a black hole at the centre of a dark matter halo. Our results indicate two phases of accretion. The first is an initial phase with a low accretion rate that depends on the black hole mass, followed by a second phase with a high accretion rate that depends on the halo mass. In the second phase, the black hole mass increases rapidly to supermassive scales, which explains the existence of quasars at redshift $z\ge 6$ and also the recently detected supermassive black holes (SMBHs) by the James Webb Space Telescope (JWST). Further, we calculate the evolution of the Eddington ratio for growing black holes. The accretion is mostly sub-Eddington except for a short super-Eddington episode when the mass accretion rate transitions from low to high. However, during that episode, the black hole mass is likely inadequate to hinder accretion through radiative feedback.
\end{abstract}




\section{Introduction}
 Black holes are ubiquitous in the Universe. A plethora of stellar black holes dot galaxies, and they formed as an end state of massive stars. However, another class of black holes occupy the centres of galaxies with masses at least a million times that of the Sun and are known as the supermassive black holes (SMBHs) \citep[e.g.][]{Kormendy95, Richstone18}. The SMBHs at low redshifts can be explained by the accretion of matter onto a seed black hole over the lifetime of a galaxy \citep[e.g.][]{Soltan82, Yu02, Shankar04}. However, the existence of supermassive black holes in the early Universe at redshifts similar to when the first galaxies originated \citep{Mortlock11, Larson23} poses a challenge to theoretical models \citep[e.g.][]{Volonteri21}.  

The black holes are known to feed on the matter in their surrounding; therefore, the accretion of matter is the popular explanation for the existence of SMBHs \citep[e.g.][]{Lynden69, Rees84}. However, the accreting gas is a source of radiation that may hinder the accretion process. A critical parameter in this regard is the Eddington ratio, which is the ratio of the force due to radiation to that due to gravitation. The accretion is called sub-Eddington for an Eddington ratio less than one and super-Eddington otherwise. A sub-Eddington accretion is adequate to explain most of the SMBHs in the present-day Universe; however, a super-Eddington accretion is required to explain the  SMBHs at high redshift \citep[e.g.][]{Haiman01, Bromm03, Regan19, Smith19}. 

Although the super-Eddington accretion is observed in some X-ray binaries \citep[e.g.][]{Okuda02} and some narrow-line Seyfert-I galaxies \citep{Mineshige00, Collin04, Du14}; however the consensus is that the super-Eddington accretion is prone to radiation feedback and outflows, hence it may not be sustained. Therefore, alternative mechanisms have been proposed to explain the high-z SMBHs. These range from SMBHs beginning growth from primordial black holes \citep[e.g.][]{Kawasaki12} to mechanisms invoking direct collapse black holes \citep[e.g.][]{Bromm03}. Some studies indicate that super-Eddington accretion may operate in brief episodes \citep{Johnson07, Park11, Tang19} or that the outflows might not be triggered until the black hole reaches a critical mass \citep[e.g.][]{King03, Li12}. The recent detection of SMBHs at very high redshifts (e.g. at  $z \approx  8.7$ by \citealt{Larson23}) has brought this debate to the forefront (see also \citealt{Maiolino1_23, Maiolino2_23, Kokorev23}). The exact mechanism of the accretion in the vicinity of a black hole, the resulting production of radiation, and its coupling with the accretion flow need to be understood.

\cite{Bondi1952} provided the analytical solution for a steady spherically symmetric flow onto a point mass. However, the current viewpoint is that the accretion flow proceeds through an accretion disc in the vicinity of the hole \citep[e.g.][]{Pringle81, Abram13}. The disc forms near the horizon of the black hole when the inflowing gas has a non-zero angular momentum. Some of the leading theories for accretion discs differ in the physical and optical thickness of the envisaged disc \citep{Narayan1998, Yuan14, Park2017}. Nonetheless, accretion disc solutions have found widespread applicability to the plethora of stellar black holes in X-ray binary systems.

In SMBHs as well, the disc likely regulates the efficiency of the radiation feedback and, hence, the accretion rate into the hole. The efficiency is hotly debated. According to some studies, the value is around $0.1$ or lower due to the photon trapping in the inner disc \citep{Begelman79}. Hence, considering that roughly $10$\% of the accretion energy is radiated \citep{Rees84, Yu02, Ueda03}, and if the accretion rate is high enough, this radiation may stop the accretion and drive outflows. However, for outflows to occur, a minimum black hole mass is required \citep{Silk98, King03}.

Cosmological simulations have focused on the SMBHs extensively \citep{Springel05, Robertson06, Hopkins08, Okamoto08, Croft09, Booth2009, McAlpine18}. A primary objective has been to reproduce the observed co-evolution of supermassive black holes and galaxies, specifically the spheroidal components of galaxies \citep{Magorrian98, Maclure02, Haring04}. Furthermore, the simulations also investigate the reported possible co-evolution of SMBH and dark matter haloes \citep{Tremaine02, Ferrarese02, Booth10, Kormendy13}. The dark matter halo is a dominant source of gravity and controls the flow of matter in a galactic system. Whether the gas that infalls into a halo finds its way to the central black hole may be challenging to model as the galaxy develops in a halo \citep[e.g.][]{Birnboim03, Dekel09}. However, in the earliest stages, the dark matter halo likely controlled the accretion rate to the hole \citep[e.g.][]{Booth10, Hobbs2012}.

Cosmological simulations often deploy a steady spherical Bondi accretion rate to model the growth of supermassive black holes \citep[e.g.][]{Springel05, Hopkins08, Booth2009, Hobbs2012}. Bondi accretion rate is popular as it captures the basic properties of the flow in a straightforward manner for studying the growth and evolution of black holes in evolving galaxies, albeit the simulations have been flexible in the actual magnitude of accretion. Often, an order of magnitude higher value is used to compensate for the artificial losses due to the lack of resolution to capture relevant subgrid physics \citep[e.g.][]{Springel05, Hopkins08}. Nevertheless, recent simulations adopt a better strategy by coupling the accretion feedback to the thermal energy of surrounding gas particles \citep{Booth2009, Schaye15}.

The standard Bondi's mass accretion rate used in cosmological simulation is for accretion without angular momentum. However, the accretion likely has angular momentum, forming a disc near the black hole that regulates the mass inflow rate and the growth of the black hole \citep[e.g.][]{Rosas15}. Another uncertainty is about the fraction of mass inflow entering the halo that finds its way to the hole and the fraction that results in the formation of the galaxy and its components \citep[e.g.][]{McAlpine17, Bower17}. A primary question in this regard is how the Bondi accretion solution and the mass accretion rate formula are modified when considering a steady spherical flow into the combined potential of the hole and the halo. The subsequent task is to put this in the context of the gas virialization and cooling in a halo and the formation of galaxy in-situ with SMBH \citep{Dubois15, Rosas16, Angles17, Habouzit17, Bower17, McAlpine17, McAlpine18, Weinberger18}. 

 \cite{Ciotti17} considered a steady isothermal accretion on a black hole in Hernquist potential \citep{Hernquist90}, and then later \cite{Ciotti18} considered the accretion in a Jaffe potential \citep{Jaffe1983}. Whether a theoretically valid hydrodynamic solution is possible for the potential of a realistic Navarro-Frenk-White (NFW) halo \citep{Navarro97} combined with that of a black hole needs to be investigated. Another related problem is the evolution of the accretion flow itself, as the seed black hole and the halo, both grow in mass with decreasing redshift. 

In this paper, we address the problem of accretion onto a black hole at the centre of a dark matter halo. We consider both a Hernquist and an NFW halo profile. We focus on bringing out the physics of an isothermal steady flow in the combined potential of the black hole and the halo. We derive the critical points of the flow, determine the possibility of trans-sonic accretion, and investigate the role of the black hole. 

We also study the consequences of our solutions for the growth of SMBHs in early haloes and then derive the evolution of the Eddington ratio to investigate the need for super-Eddington accretion to explain the high redshift SMBHs detected by JWST and earlier studies \citep{Fan01, Fan03, Mortlock11, Inayoshi20, Larson23}.  

The paper is organised as follows. In section 2, we present the basic equations for the flow. We provide the solutions for a Hernquist dark matter halo without and with a seed black hole. In section 3, we investigate the solutions for an NFW halo and derive the mass accretion rate from the knowledge of the critical points of the flow. In section 4, we calculate the redshift evolution of the black hole mass as a function of redshift, derive the Eddington ratio for the growing black hole, and compare these results with observations. We discuss and summarise our findings in section 5.

\section{Spherically symmetric accretion into dark matter haloes}
The equations for the mass and the momentum conservation for a spherically symmetric steady flow are,
\begin{equation}
 \frac{d}{dr} (\rho v r^2) = 0 \label{mass}
\end{equation}
\begin{equation}
    v\frac{dv}{dr} = -\frac{1}{\rho}\frac{dp}{dr} - \frac{d\phi}{dr} \label{moment}
\end{equation}
where $\rho$ is the density, $p$ is the pressure, $v$ is the velocity and $\phi$ is the gravitational potential.  
With sound speed $c_s = \sqrt{dp/d\rho} $, and a dimensionless Mach number, $\mathcal{M}=v/c_s$, the above equations can be transformed to,   
\begin{equation}
    \frac{\mathcal{M}^2-1}{2c_s^2}\frac{dc_s^2}{dr} + \frac{\mathcal{M}^2-1}{2\mathcal{M}^2}\frac{d\mathcal{M}^2}{dr}= \frac{2}{r} - \frac{1}{c_s^2} \frac{d\phi}{dr} \ .
\label{flow_eq}
\end{equation}
The sound speed, $c_s = \sqrt{{\rm k_B} T / (\mu {\rm m_H}})$, where $\mu$ is the mean molecular weight, $m_{\rm H}$ is the mass of the hydrogen atom, and $T$ is the temperature which is constant for an isothermal flow. Therefore Eq~\ref{flow_eq} can be readily solved to obtain,
\begin{equation}
\frac{\mathcal{M}^2}{2} - \ln{\mathcal{M}} = 2 \ln{r} - \frac{\phi}{c_s^2} + {\rm const.} 
\label{solution} 
\end{equation}

\subsection{Accretion into the centre of a dark matter halo}
In this section, we consider an isothermal accretion flow into a dark matter halo. We first consider the Hernquist profile for the halo  \citep{Hernquist90}.
 The potential for a Hernquist dark matter halo is, 
\begin{equation}
    \phi = -\frac{G M_{\rm h}}{r+a}
\end{equation}
where $a$ is a scale radius that encloses 25\% of the halo mass, $M(a) = M_{\rm h}/4$. The Hernquist profile features a shallower decrease in density with radial distance in the inner halo ($r< a$), and a rapid decrease in the outer halo, similar to the NFW  profile.
From Eq.~\ref{flow_eq}, for an isothermal flow into a Hernquist potential,
\begin{equation}
    \frac{\mach^2 - 1}{ \mach} \frac{d \mach}{d r} =  \frac{2}{r} - \frac{G M_{\rm h}}{c_{\rm s}^2 (r+a)^2} \ .
    \label{eq_iso_flow}
\end{equation}
The solution is, 
\begin{equation}
    \frac{\mathcal{M}^2}{2} - \ln{\mathcal{M}} = 2 \ln{\frac{r}{l_{\rm h}}} + \frac{l_{\rm h}}{r + a} + C \ ,
    \label{eq_family}
\end{equation}
where  $l_{\rm h} = G\, M_{\rm h}/c_s^2$, and $C$ is a constant. 
\begin{figure}
    \centering
    \includegraphics[width = \linewidth]{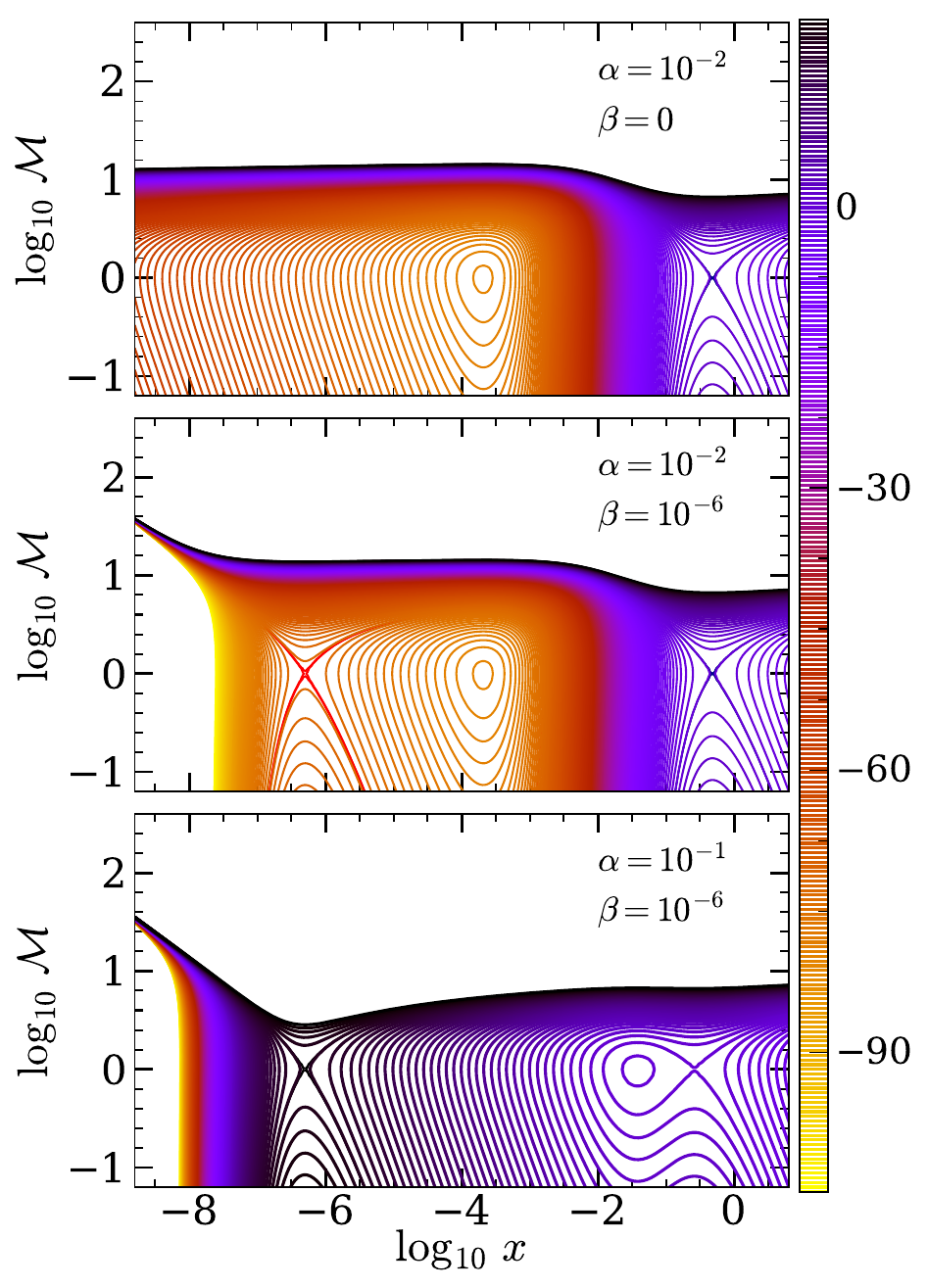}
    \caption{The family of solutions for the steady spherical flow to the centre of a Hernquist dark matter halo, obtained by varying $C$ in Eq.~\ref{eq_family}; showing Mach number, $\mathcal{M}$, as a function of the dimensionless radial distance, $x=r/l_{\rm h}$,  where $l_h = G M_{\rm h}/c_{\rm s}^2$. Top panel: if no black hole is present at the centre ($\beta=M_{\rm bh}/M_{\rm h}=0$), there are two critical points, and the trans-sonic flow does not go all the way to the centre. If a black hole is present, then there are three critical points, and the trans-sonic flow into the centre is possible either through the innermost critical point (bottom panel) or through the outermost critical point (middle panel) depending on the value of $\alpha=a/l_{\rm h}$. The bottom panel is for $\beta=10^{-6}$ and $\alpha=10^{-1}$.  The middle panel is for the same $\beta=10^{-6}$ but for a lower $\alpha=10^{-2}$. The location of the critical points depends on the values of $\alpha$ and $\beta$, and can also be visualised in Fig.~\ref{fig_critical}.}
    \label{fig_M_x}
\end{figure}
The family of solutions is shown in Fig.~\ref{fig_M_x} top panel by varying the value of $C$. The flow can be analysed by the knowledge of critical points. The location of the critical points of the flow ($r_c$) can be determined by setting the right-hand side of Eq.~\ref{eq_iso_flow} to zero, to obtain,
\begin{equation}
    x_c = \frac{1}{4} \left[ (1 - 4 \alpha) \pm \sqrt{1 - 8 \alpha} \right]\ ,
\end{equation}
 where, $x_c = r_c/l_h$,  and,   $\alpha = a/l_{\rm h}$. Two critical points are possible for $\alpha < 1/8$, one is located at $r_c < a$, the other at $r_c > a$ (see red curve in Fig.~\ref{fig_critical}). The two critical points approach each other as  $a$ increases or $l_{\rm h}$ decreases, and for $\alpha=a/l_{\rm h}=1/8$, they merge into a single critical point at $r_c=a$. No critical point exists for $\alpha>1/8$. 

 The condition, $\alpha\le 1/8$, for the existence of critical points can be translated to the ratio of a characteristic halo velocity and the sound speed, $ v_{a} \ge 2\sqrt{2} c_s$, where $v_{a}^2=G M_{\rm h}/a$. Alternatively, the same condition can be translated to a maximum flow temperature, $T_{\rm max}$ corresponding to a specific halo mass, such that $T < T_{\rm max}$ for trans-sonic accretion, and,
 \begin{equation}
 T_{\rm max} = \frac{GM_{\rm h} \mu m_{\rm H}}{8 k_{\rm B} a} \approx 10^6~{\rm K}  \left[\frac{M_{\rm h}}{10^{12}~{\rm M}_\odot}\right] \left[\frac{a}{\rm 40~kpc}\right]^{-1} \ ,
 \end{equation}
 where we have used $\mu=0.6$. We note that this maximum temperature is roughly the same as the virial temperature of the halo. An analogous calculation for an NFW halo with concentration parameter $c \approx 5$ would yield $T_{\rm max}\sim T_{\rm vir}$ with $\alpha$ in that case is defined as $\alpha = R_{\rm s }/l_{\rm h}$ where $R_s$ is the NFW scale radius related to the virial radius as $R_{s}=R_{\rm vir}/c$.

 For  $\alpha$ less than a critical value, which is $1/8$ for the Hernquist halo (and approximately $0.1$ for the NFW halo with  $c = 5$), the flow can proceed along the solution through the outer critical point; however, such a solution is unphysical as it loops around the inner critical point and is double valued  (Fig.~\ref{fig_M_x} top panel). Therefore, trans-sonic accretion to the centre of a Hernquist or an NFW dark matter halo is impossible. However, this outcome changes when a seed black hole is present as we investigate in the next section.

\subsection{Accretion onto a seed black hole in a Hernquist dark matter halo}
\begin{figure}
    \centering
    \includegraphics[width=\linewidth]{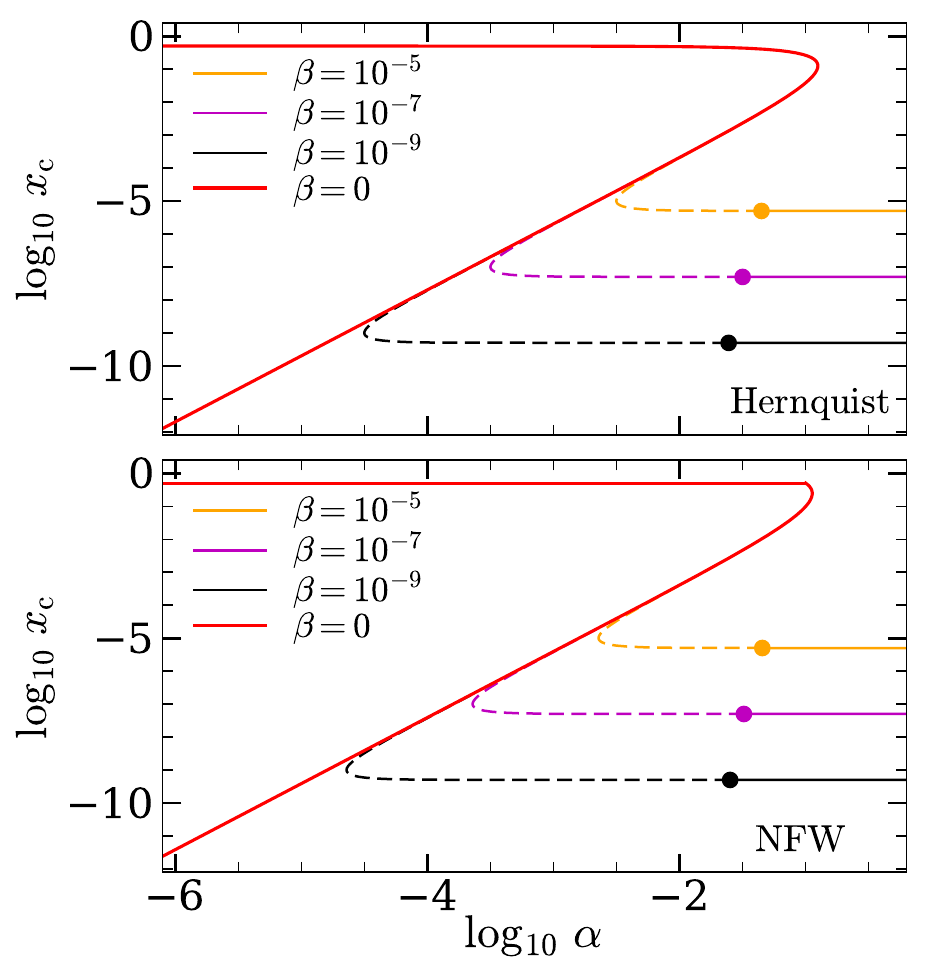}
    \caption{  
    Top panel: Location of the critical points for accretion into the combined potential of a black hole and a Hernquist halo, shown as a function of $\alpha=a/l_{\rm h}$, for $\beta=0$ (red), $10^{-9}$ (black), $10^{-7}$ (magenta) and $10^{-5}$ (yellow). The critical points are the points at which the vertical lines corresponding to a constant $\alpha$ intersect the curves corresponding to a particular $\beta$. For $\beta=0$, there are a maximum of two critical points, and for $\beta \neq 0$, a maximum of three critical points. Further, for non-zero $\beta$, the trans-sonic flow to the centre may proceed through the innermost or the outermost critical point. For high $\alpha$, the solution proceeds through the innermost critical point; however, as $\alpha$ decreases below a transition value (shown as filled circles) which depends on $\beta$ (Eq.~\ref{eq_turnover} for Hernquist and Eq.~\ref{eq_turnover_nfw} for NFW), the solution through the innermost critical point is unphysical (shown by the dashed portion of the curves and also see the red curve in the middle panel of Fig.~\ref{fig_M_x}), then the other trans-sonic solution through the outermost critical point takes-over. Bottom panel: same as the top panel but for an NFW halo potential with concentration parameter $c=5$.} 
    \label{fig_critical}
\end{figure}

In this section, we consider the effect of the presence of a seed black hole at the centre of a Hernquist halo. The combined potential of a Hernquist dark matter halo and the black hole at its centre is, 
\begin{equation}
    \phi = - \frac{G M_{\rm h}}{r+a} - \frac{G M_{\rm bh}}{r} = -G M_{\rm h} \left[ \frac{1}{r+a} + \frac{\beta}{r}\right]
    \label{eq_comb_pot_hern}
\end{equation}
where we have defined the ratio of the black hole mass to the halo as $\beta = M_{\rm bh}/M_{\rm h}$. The flow solution for this total potential is,
\begin{equation}
    \frac{\mathcal{M}^2}{2} - \ln{\mathcal{M}} = 2 \ln{\frac{r}{l_{\rm h}}} + \frac{l_{\rm h}}{r + a} + \beta \frac{l_{\rm h}}{r} + C
    \label{eq_hern_bh_sol}
\end{equation}
The critical points in this case are given by,
\begin{equation}
    \frac{2}{r_c} =  \frac{G M_{\rm h}}{c_s^2}\Big[\frac{1}{(r_c+a)^2} + \frac{\beta}{r_c^2} \Big]
\end{equation}
By defining the dimensionless parameters, $ l_{\rm h} = {\rm G}M_{\rm h}/c_{\rm s}^2$, $x=r/l_{\rm h}$, $x_{\rm c} = r_{\rm c}/l_{\rm h}$ and $\alpha = a/l_{\rm h}$, the following equation is obtained for the critical points,
\begin{equation}
    2 x_{\rm c} (x_{\rm c} + \alpha)^2 ~=~ x_{\rm c}^2 + \beta (x_{\rm c} + \alpha)^2  \label{eq_cubic}
\end{equation}
In Fig.~\ref{fig_critical}, we have shown the locus of critical points as a function of $\alpha$ for four different values of $\beta$. The location of the critical point for a specific $\alpha$ and $\beta$ can be read from the appropriate curve. 

In general, there are three critical points for a specific $\alpha$ when a black hole is present (black, magenta, and orange curves) compared to only two when there is no black hole (red curve). A trans-sonic flow into the black hole can proceed through the innermost or the outermost critical point.

The trans-sonic flow through the innermost critical point is shown in Fig.~\ref{fig_M_x}  bottom panel, and this solution is akin to the standard Bondi accretion onto a black hole. The trans-sonic flow through the outermost critical point (in Fig.~\ref{fig_M_x} middle panel) is due to the halo. There are unphysical solutions as well, shown in Fig.~\ref{fig_M_x}. For example, in the middle panel, the solution shown by the red curve passes through the innermost critical point, but then it loops around the intermediate critical point and turns back; hence, it is double-valued and unphysical. The corresponding values of $\alpha$ and $\beta$ for which such unphysical solutions occur are shown as dashed part of the curves in Fig.~\ref{fig_critical}). 

The two types of solutions operate in two separate regimes of $\alpha$. Also, the mass accretion rates for the two types of solutions are significantly different. The mass accretion rate can be calculated from Eq.~\ref{mass} and \ref{moment}; then using  $v=c_s$ at the critical point $r=r_c$; and $\rho=\rho_{\infty}$, $v=0$ at infinity. The mass accretion rate is thus given by,
\begin{equation}
    \dot M_{\rm bh} = \frac{4\pi {\rm G^2} M_{\rm h}^2\rho_\infty}{c_{\rm s}^3} x_c^2 e^{\left[-\frac{\phi}{c_s^2} - \frac{1}{2} \right]_{x=x_c} }
    \label{eq_Macc_phi}
\end{equation}
For the combined potential of a Hernquist dark matter halo and a black hole (Eq.~\ref{eq_comb_pot_hern}), we obtain, 
\begin{eqnarray}
     \dot M_{\rm bh} &=& \frac{4\pi {\rm G^2} M_{\rm h}^2\rho_\infty}{c_{\rm s}^3} \lambda \nonumber \ ;\\
     \lambda &=& x_c^2 e^{\left[\frac{1}{x_c+\alpha} +\frac{\beta}{x_c} - \frac{1}{2} \right] }    
    \label{eq_Macc_xc}
\end{eqnarray}
where $\lambda$ is the dimensionless mass accretion rate. If the location of the critical point is known, we can obtain the exact value of the mass accretion rate. The innermost and outermost critical points can be recovered analytically. 

The innermost critical point can be recovered by setting the limit $x_c\ll \alpha$ in Eq.~\ref{eq_cubic}, which yields $x_c\approx \beta/2$. For this, the dimensionless mass accretion rate (from Eq.~\ref{eq_Macc_xc}) is given by, 
\begin{equation}
     \lambda =  \frac{\beta^2}{4}   e^{\left[\frac{2}{\beta+2\alpha} + \frac{3}{2}\right]} \ . 
    \label{eq_mass_acc_hern}
\end{equation}
For high values of $\alpha$, $\lambda \approx 1.12 \beta^2$, the same as for the standard Bondi accretion, which is not surprising since the innermost critical point is due to the black hole, and the flow through it is akin to the standard Bondi accretion on to a point mass.

The outermost critical point is due to the halo, can also be recovered analytically, in the limit $ x_{\rm c} \gg \alpha$. Then Eq.~\ref{eq_cubic} can be written as,
\begin{equation}
    2 x_{\rm c}\big[1 + \alpha/x_{\rm c} \big]^2 - \beta \big[1 + \alpha/x_{\rm c} \big]^2 = 1 \ .
    \label{eq_crit_simple}
\end{equation}
The critical point is then given by,
$    x_{\rm c} \approx \big(1 + \beta \big)/2 $.
Substituting the critical point in Eq.~\ref{eq_Macc_xc} together with $\alpha/x_c \approx 0$ yields the following expression for the dimensionless mass accretion rate,
\begin{equation}
     \lambda =  (1  + \beta)^2\frac{ e^{3/2}}{4}   \ . 
    \label{eq_mass_acc_hern2}
\end{equation}
As $\beta\ll 1$, we have $\lambda \approx 1.12$, which also seems similar to the standard Bondi accretion; however, there is a subtle difference that the mass accretion rate is proportional to $M_{\rm h}^2$ instead of $M_{\rm bh}^2$ (see Eq.~\ref{eq_Macc_xc}).

The two trans-sonic solutions are separated by a transition $\alpha$, and for the transition $\alpha$, the flow solution passes through both the innermost and the outermost critical point.  

\begin{figure}
    \centering
    \includegraphics[width= \linewidth]{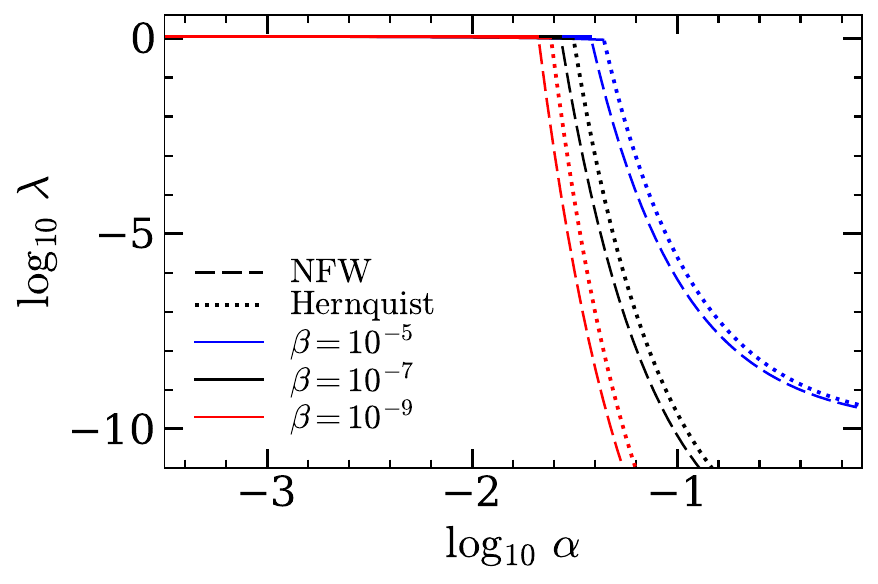}
    \caption{The dimensionless mass accretion rate, $\lambda$, as a function of $\alpha$, for $\beta = 10^{-9}$ (red), $\beta = 10^{-7}$ (black) and $\beta = 10^{-5}$ (blue); for a Hernquist halo (dotted curves), and an NFW halo (dashed curve). Also, the parameter $\alpha=a/l_{\rm h}$ for the Hernquist halo, and $\alpha=R_s/l_{\rm h}$ for the NFW halo, where $R_s$ is the NFW scale radius, related to the virial radius as $R_s=R_{\rm vir}/c$, and we take $c=5$. 
    The dashed portion of the curves shows the mass accretion rate when the trans-sonic solution is through the innermost critical point (that occurs for a higher $\alpha$). The solid portion of the curve shows the mass accretion rate when the flow proceeds through the outermost critical point (for smaller $\alpha$). $\lambda$ increases from a low value ($\approx \beta^2$) with decreasing $\alpha$ and becomes constant ($\lambda\approx 1.12$) when the trans-sonic solution through the outermost critical point takes over.}
    \label{fig_Mdot_alpha}
\end{figure}

The transition $\alpha$ depends on the value of $\beta$ and can be derived by imposing the condition that the trans-sonic flow passes through both the outermost and the innermost critical point (see Appendix~\ref{sec_app}). The transition $\alpha$ thus recovered is,
\begin{equation}
    \alpha_{\rm T} \approx \frac{1}{2 \ln (1/\beta) }\ .
    \label{eq_turnover}
\end{equation}
For $\alpha \gg \alpha_{\rm T}$, the mass accretion rate is low, and it approaches the standard Bondi's mass accretion rate \citep{Bondi1952}. For $\alpha \ll \alpha_{\rm T}$, the mass accretion rate becomes high ($\lambda \approx 1.12$ or $\dot M_{\rm bh} \propto M_{\rm h}^2$) as it is controlled by the halo (Fig.~\ref{fig_Mdot_alpha}). 

The value of $\alpha$ decreases with redshift as the halo mass increases \citep{Correa1}, indicating two phases of accretion, beginning with a phase of slow growth initially for a high $\alpha \gg \alpha_{\rm T}$, and then a phase of rapid growth at $\alpha$ below $\alpha_{\rm T}$ (Eq.~\ref{eq_turnover}).

\section{Accretion onto a seed black hole in an NFW dark matter halo}

\begin{figure}
    \centering
    \includegraphics[width= \linewidth]{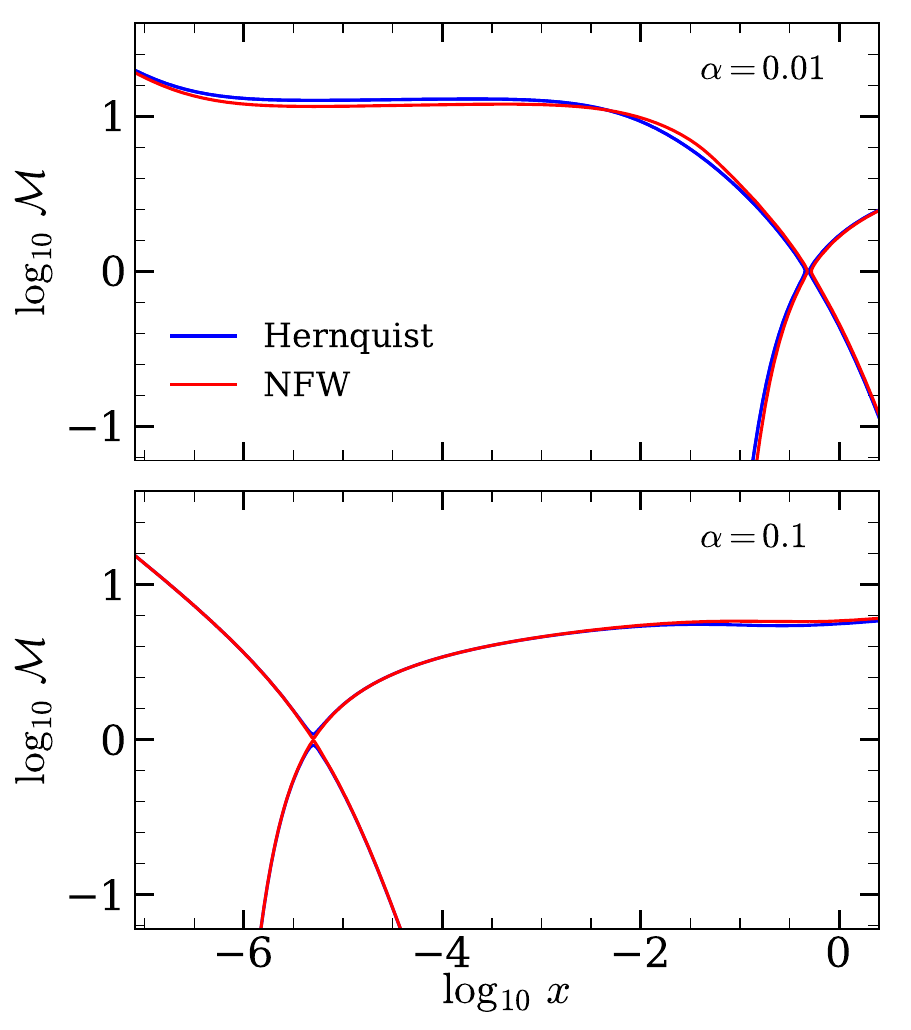}
    \caption{The Mach number, $\mathcal{M}$, as a function of $x=r/l_{\rm h}$ for accretion onto the black hole residing in dark matter halo shown by the curves that rise towards the centre; for a Hernquist halo  (blue curves) and an NFW halo (red curves). For NFW haloes, we have chosen a concentration parameter, $c=5$, that is appropriate for high redshift haloes \citep[e.g.][]{Prada12, Correa3}. In general, there are three critical points in each case. Top panel: for the trans-sonic flow through the outermost critical point (that occurs for $\alpha=0.01,\ \beta = 10^{-6}$). Bottom panel: for the flow through the innermost critical point (for $\alpha=0.1$ and  $\beta=10^{-6}$).   For the Hernquist halo $\alpha=a/l_h$, and for the NFW halo, $\alpha = R_{\rm s}/l_{\rm h}$,  where $R_{\rm s}$ is the scale-radius that is related to the virial (truncation) radius as $R_{\rm s} = R_{\rm vir}/c$. The paramter $l_{\rm h} = G M_{\rm h}/c_s^2$.}
    \label{fig_nfw_hern}
\end{figure}

In this section, we investigate the flow onto a black hole in an NFW halo \citep{Navarro97}. The NFW is a universal halo profile obtained for haloes in cosmological simulations of galaxy formation. The potential for the NFW profile is, 
\begin{equation}
    \phi_{\rm nfw} = -\frac{ G M_{\rm h}}{f(c)}\frac{\ln(1+r/R_s)}{r} + \frac{GM_{\rm h}}{R_{\rm vir}f(c)} \frac{c}{1+c}
    \label{eq_phinfw}
\end{equation}
where $M_{\rm h}$ is the halo mass within a virial radius, $R_{\rm vir}$.  $f(c) = \ln(1+c) - c/(1+c)$, and $R_s$ is the scale radius that is related to $R_{\rm vir}$ through the concentration parameter ($c$) as $R_s = R_{\rm vir}/c$.  The scale-radius is the radial distance at which the density profile changes its slope from a shallow $r^{-1}$ variation to a steep $r^{-3}$ variation. The equivalent parameter in a Hernquist halo is $a$. Equation~\ref{eq_phinfw} is valid for $r\le R_{\rm vir}$, after which the halo is truncated and $\phi_{\rm nfw } \propto 1/r$.

In a Hernquist halo, the mass remains finite when the density profile is integrated up to infinity. However, for the NFW halo, the mass diverges when integrating the density profile to infinity. Therefore, the halo's mass is defined up to a virial radius at which the halo profile is truncated. The virial radius is defined as \citep[e.g.][]{Mo1998}, 
\begin{equation}
R_{\rm vir} = \left[\frac{3 M_{\rm h}}{4 \pi (200 \rho_{\rm crit})}\right]^{1/3}  \ .
\end{equation}

For the NFW potential combined with the potential of a black hole at its centre, following the method in the previous section, the flow solution is given by,
\begin{equation}
    \frac{\mathcal{M}^2}{2} - \ln{\mathcal{M}} = 2 \ln(x) + \frac{\ln(1+x/\alpha)}{f(c) x} + \frac{\beta}{x} + C \ ,
\end{equation}
where $x=r/l_{\rm h}$ with $l_{\rm h}=G M_{\rm h}/c_s^2$ and this time $\alpha=R_s/l_{\rm h}$ as it features the scale radius $R_s$ of an NFW halo instead of the parameter $a$ of a Hernquist halo. The critical points are given by,
\begin{equation}
\frac{2}{x_c} =  \frac{\ln(1+x_c/\alpha) - \frac{x_c/\alpha}{1+x_c/\alpha}}{x_c^2 f(c)} + \frac{\beta}{x_c^2}  \ .
\label{eq_crit_nfw}
\end{equation}
Locations of the critical points for specific values of $\alpha$ and $\beta$ are shown in Fig.~\ref{fig_critical} bottom panel. There are three critical points: innermost, intermediate, and outermost.

The innermost critical point can be recovered in the limit $x_c\ll\alpha$. Therefore, setting $x_c/\alpha\approx 0$ in Eq.~\ref{eq_crit_nfw} gives,
$x_c\approx\beta/2$, same as what we obtained for the Hernquist halo. 

The outermost critical point is difficult to retrieve analytically. However, we can use the fact that it is approximately at the virial radius or outside it (Fig.~\ref{fig_critical}, see also \citealt{Hobbs2012}), and the NFW halo is truncated at the virial radius. Therefore, the first term on the right-hand side of Eq.~\ref{eq_crit_nfw} is $1/x_c^2$, which yields $x_c\approx (1+\beta)/2$, same as in the case of the Hernquist halo. 

In Fig.~\ref{fig_nfw_hern}, we show a comparison of the trans-sonic solutions for a Hernquist halo and an NFW halo (for $c=5$).  
The two types of trans-sonic solutions are shown: the solution through the innermost critical point (bottom panel) and through the outermost critical point (top panel). Solutions through the innermost point for the two types of dark matter haloes are almost identical because the innermost point is due to the black hole. Moreover, the solutions through the outermost critical point differ only slightly between the two haloes.

The solutions for an NFW halo depend on the concentration parameter $c$. However, our results address the high redshift haloes at $z\geq 6$, for which the concentration parameter varies mildly ($c\approx 5$) \citep[e.g.][]{Prada12,Correa3}.

For both the Hernquist and the NFW halo, there are three critical points: innermost, intermediate, and outermost (Fig.~\ref{fig_critical}). The Mach number, $\mathcal{M}$, for the accretion flow through the outermost critical point (top panel of Fig.~\ref{fig_nfw_hern}), stops increasing as $x$ is closer to the location of the intermediate critical point. Then, there are three possibilities depending on the value of $\alpha$. First, for very low $\alpha$, $\mathcal{M}$ increases again as $x$ approaches the location of the innermost critical point. Second, when $\alpha$ is higher and equal to a transition value ($\alpha_{\rm T}$), $\mathcal{M}$ decreases and passes through the innermost critical point. Third, for very high $\alpha$, $\mathcal{M}$ decreases rapidly, and the solution cannot reach the innermost critical point; instead, it loops around the intermediate critical point. 

In the third case, the solution through the outermost critical point cannot reach the black hole. However, the accretion occurs as the trans-sonic solution through the innermost critical point takes over (Fig.~\ref{fig_nfw_hern} bottom panel). 

The value of $\alpha$ decreases with redshift, and we can also understand the switch-over between the two solutions from the perspective of a decreasing $\alpha$. For a very high value of $\alpha$, the feasible trans-sonic solution for the growth of the black hole is through the innermost critical point. As $\alpha$ decreases below the transition value, $\alpha_{\rm T}$, the solution through the outermost critical point takes over. At $\alpha=\alpha_{\rm T}$, the two types of solutions merge. 

The transition value $\alpha_{\rm T}$ depends on the value of $\beta$ and can be derived by imposing the condition that the trans-sonic flow passes through both the innermost and the outermost critical point. The transition $\alpha$ then is,
\begin{equation}
    \alpha_{\rm T} \approx \frac{1}{2 \eta \ln (1/\beta)}\ .
    \label{eq_turnover_nfw}
\end{equation}
where $\eta = (1+1/c)\ln(1+c) - 1$. The transition $\alpha$ is shown as a dashed curve in Fig.~\ref{fig_Mdot}. For a seed black hole of $10$~M$_\odot$, and a halo mass $10^{12}$~M$_{\odot}$, the transition $\alpha$ is  $\approx 0.02$. The transition from one solution to the other leads to two distinct phases of black hole growth, as we discuss in the next section.

\subsection{The black hole mass accretion rate in an NFW halo}

\begin{figure}
    \centering
    \includegraphics[width= \linewidth]{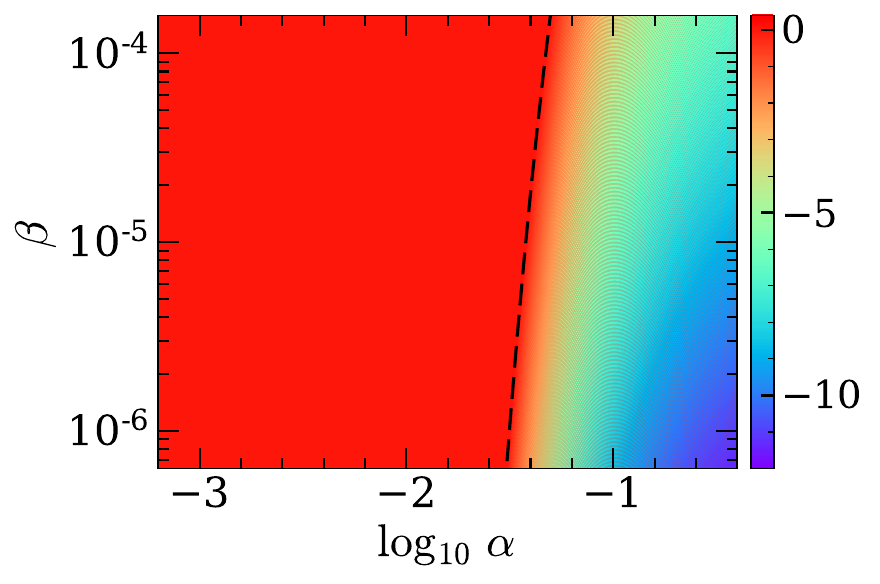}
    \caption{The dimensionless mass accretion rate onto a black hole at the centre of an NFW halo, $\log_{10}\lambda$ (Eq.~\ref{eq_lamb_nfw}) shown in colour as a function of $\alpha=R_s/l_h$ and $\beta = M_{\rm bh}/M_{\rm h}$, where $R_s$ is the NFW scale radius related to the virial radius through a concentration parameter $c$ as $R_s=R_{\rm vir}/c$, and we take $c=5$. The mass accretion rate is low, proportional to $\beta^{2}$ for high $\alpha$, and it is constant ($\approx 1.12$) for low values of $\alpha$. The two regimes (phases) are separated by a transition $\alpha$ (Eq.~\ref{eq_turnover_nfw}) shown as a dashed curve.}
    \label{fig_Mdot}
\end{figure}

The black hole mass accretion rate for the two solutions through the innermost or outermost critical points can be obtained by using Eq.~\ref{eq_Macc_phi}, with the potential $\phi$ this time representing the combined potential of an NFW halo and a black hole. The mass accretion rate thus obtained is, 
\begin{eqnarray}
     \dot M_{\rm bh} = \frac{4\pi {\rm G^2} M_{\rm h}^2\rho_\infty}{c_{\rm s}^3} \lambda \ ; \qquad \qquad  \qquad \qquad \qquad   \nonumber \\
     \lambda =      
     \begin{cases}
x_c^2 e^{\left[ \frac{\ln(1+{x_c / \alpha})}{x_c  f(c)} - \frac{1/(1+c)}{\alpha f(c)}  +  \frac{\beta}{x_c} - \frac{1}{2}     \right]}  &\text{$x_c \ll \alpha$}
\\ 
x_c^2 e^{\left[ \frac{1+\beta}{x_c}  - \frac{1}{2}     \right]}    &\text{$x_c \gg \alpha$} 
\end{cases} 
\label{eq_mass_acc_nfw}
\end{eqnarray}
By substituting $x_c=\beta/2$ for the innermost critical point and $x_c \approx (1+\beta)/2$ for the outermost critical point that is outside the virial radius, we retrieve the following expression revealing two phases of mass accretion:
\begin{equation}
\lambda \approx \begin{cases}
\frac{\beta^2}{4} e^\frac{c/(1+c)}{\alpha f(c)}  e^{3/2} \approx 1.12 \beta^2 e^\frac{c/(1+c)}{\alpha f(c)}&\text{$\alpha \gg \alpha_{\rm T}$}\\
\frac{(1+\beta)^2}{4} e^{3/2} \approx 1.12 &\text{$\alpha \ll \alpha_{\rm T}$}
\end{cases}
\label{eq_lamb_nfw}
\end{equation}
The dimensionless mass accretion rate, $\lambda$, is plotted in Fig.~\ref{fig_Mdot_alpha} and Fig.~\ref{fig_Mdot} in the parameter space of $\alpha$ and $\beta$.
The initial phase  (for $\alpha\gg \alpha_{\rm T}$) exhibits a low mass accretion rate, and the black hole mass remains almost the same as the
initial seed mass. The initial `slow' phase of growth is set by the solution through the innermost critical point that is due to a high $\alpha$ (Fig.~\ref{fig_nfw_hern} bottom panel). As the value of $\alpha=c_s^2 R_s/(GM_{\rm h})$ decreases with redshift due to increasing halo mass, at a specific $\alpha$ (Eq.~\ref{eq_turnover_nfw} and the dashed line in Fig.~\ref{fig_Mdot}) the solution transitions to the one through the outermost critical point (Fig.~\ref{fig_nfw_hern} top panel). Then, the second phase of a `rapid' growth with a high mass accretion rate begins (flat line in Fig.~\ref{fig_Mdot_alpha} and the red zone in Fig.~\ref{fig_Mdot}). In this phase, the dimensionless mass accretion rate, $\lambda$, is constant, and the actual accretion rate is proportional to $M_{\rm h}^2$. 

The phases of evolution are also proposed heuristically by \cite{Li12}; and with cosmological simulations \cite[e.g.][]{Bower17, Dubois15, Angles17, McAlpine18}; however, the origin of phases reported by the simulations is different from ours due to two reasons. Firstly, the seed black hole mass used in the simulations is $\sim 10^5$~M$_\odot$ \citep{Dubois15, Bower17, McAlpine18}, while in our work, the black hole grows from an initial mass of $10$~M$_\odot$, typical for a stellar black hole. Secondly, the simulations tie the phases of SMBH growth with supernovae feedback \citep{Dubois15, Angles17}. \cite{Bower17} attributes the initial slow phase to the clearance of gas by `buoyant' outflows and the rapid `nonlinear' phase to the failure of outflows due to increased virial temperature of the halo that results in the availability of gas in the halo  \citep[see also][]{McAlpine17, McAlpine18}. However, in our work, the origin of the phases of black hole growth is purely hydrodynamic.

\section{Growth of the central black hole}

In this section, we study the growth of a black hole in high redshift haloes as a consequence of the accretion rate derived in the previous section. 

\begin{figure}
    \centering
    \includegraphics[width=\linewidth]{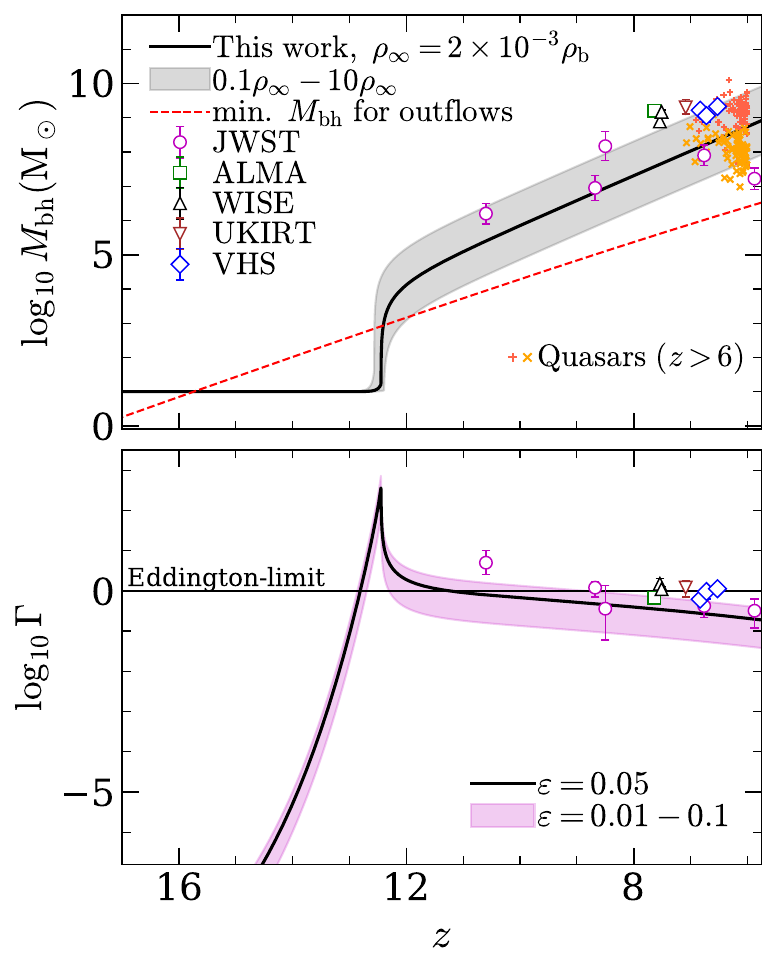}
    \caption{Top panel: mass of the growing black hole, $M_{\rm bh}$ with redshift, $z$. The results are shown for a seed black hole of mass $10$~M$_\odot$ and for an accretion flow with $T=10^4$~K and $\rho_{\infty}=2\times 10^{-3} \rho_{\rm b}$ where $\rho_{\rm b}$ is the cosmic baryon density. The shaded region shows the range when varying  $\rho_{\infty}$ by a multiplication factor of $0.1-10$. The magenta circles are for supermassive black holes (SMBHs) detected recently by the JWST \citep{Larson23, Maiolino1_23, Maiolino2_23, Kokorev23}; the green square is for the SMBH in a quasar detected by ALMA \citep{Wang21}; black triangles are for the SMBHs detected by WISE \citep{Banados18, Yang20}; red down-pointing triangle is for the SMBH mass determined from quasar observation using UKIRT \citep{Mortlock11},  blue diamonds are for the detection by VHS \citep{Reed19}. The red small plus symbols are for the quasars detected at $z \ge 6$ compiled by  \citet{Banados16}, and orange small cross symbols are for the quasars from \citet{matsuoka18a, Matsuoka18b, Matsuoka19a, Matsuoka19b}; see also \citealt{Inayoshi20} for a compilation of the quasar data. Bottom panel: Eddington ratio corresponding to the black curve in the top panel calculated for a radiative efficiency, $\varepsilon = 0.05$,  compared with the measured values from recent observations shown as datapoints with errorbars (symbols for references are the same as in the top panel). The shaded region shows the possible range if $\varepsilon$ is varied from $0.01$ to $0.1$. }
\label{fig_final}
\end{figure}

The accretion rate depends on the temperature and the density of the gas available for accretion far away. It also depends on the mass of the dark matter halo that evolves with redshift as, 
$    M_{\rm h}(z) = M_{\rm h,0} (1+z)^{a} e^{-b z} $,
where $a$ and $b$ depend on $M_{\rm h,0}$.  With this, Eq.~\ref{eq_mass_acc_nfw} can be written in the following differential equation form, 
\begin{equation}
    \frac{dM_{\rm bh}}{dt} =
    \frac{4 \pi {\rm G}^2 \rho_{\infty,0} M_{\rm h,0}^2}{c_{\rm s}^3} \lambda \big(1+z\big)^{3+2 a} e^{-2 b z} 
\end{equation}
where, $\rho_{\infty,0} = \rho_{\infty} (1+z)^{-3}$. We use $dt/dz = -(1+z)^{-1}H(z)^{-1}$, and the accretion histories from \cite{Correa1} to solve the above differential equation, with the latest cosmological parameters from \cite{Planck20}. Figure~\ref{fig_final} (top panel, solid black curve) shows the evolution of a seed black hole of $10$~M$_\odot$ in a massive halo $M_{\rm h,0}=10^{13}$~M$_\odot$, for a flow temperature $T=10^4$~K and a fiducial value of density far away, $\rho_{\infty}=2\times 10^{-3}\rho_{\rm b}$,  where $\rho_{\rm b}$ is the cosmic baryon density. The shaded region shows the range when varying the density from $0.1\rho_{\infty}$ to $10\rho_{\infty}$.
The two phases of accretion are evident (black curve), as the black hole has negligible growth up to a redshift of $\approx 12.5$, but after that, its mass increases rapidly to supermassive scales. At redshift $8$, it grows to $\approx 10^8$~M$_\odot$. The predicted evolution as shown by the black curve and the shaded region, agrees with the masses of the SMBHs detected recently by the JWST \citep{Larson23, Maiolino1_23, Maiolino2_23, Kokorev23} and from earlier detection of quasars \citep{Banados16, Banados18, Mortlock11, Reed19, matsuoka18a, Matsuoka18b, Matsuoka19a, Matsuoka19b, Yang20, Wang21, Inayoshi20}.

We also calculate the Eddington ratio for the black holes in this work (Fig.~\ref{fig_final}, bottom panel). For that, we adopt the standard relation, $
L = \varepsilon  \dot{M}_{\rm bh}  c^2
$,  to estimate the luminosity for an SMBH. Then the Eddington ratio is simply given by, $\Gamma = L/L_{\rm Edd}$, where $L_{\rm Edd} = [1.3\times 10^{38}~{\rm erg~s^{-1}}]$ $M_{\rm bh}/{\rm M}_{\odot}$. 

Figure~\ref{fig_final} (bottom panel) shows the evolution of the Eddington ratio, $\Gamma$,  as a solid black curve corresponding to the growth of a seed black hole of $10$~M$_\odot$ shown with a black curve in the top panel. We use a radiative efficiency of $\varepsilon =0.05$, and also show the effects of its variation from  $0.01$ to $0.1$ \citep[and references therein]{Inayoshi20}. 
The Eddington ratio, $\Gamma$, is low at the beginning at high redshift ($z> 13$), and then it increases to super-Eddington values for a short period of roughly $100$~Myr. This period is likely even shorter as the radiative efficiency may be lower due to photon trapping \citep{Begelman79} or magnetic-field effects \citep{Narayan03, Sadowski15, McKinney15}. 

The Eddington ratio begins decaying rapidly as the black hole enters into the rapid phase of growth at $z\approx 12.5$ (Fig.~\ref{fig_final} bottom panel). Eventually $\Gamma$ settles to sub-Eddington values,  $\approx 0.1$  at $z\approx 6$, similar to the values for quasars observed at high redshifts \citep[e.g.][]{Mortlock11, Banados16, Banados18, Yang20, Wang21, Maiolino1_23, Maiolino2_23, Kokorev23, Larson23}. 

The accretion flow may not continue if there is strong radiative feedback during the super-Eddington episode. However, the super-Eddington episode is short-lived (roughly 100 Myr), and more importantly, the black hole mass is low during that episode, which may not be adequate to drive outflows and stop accretion \citep{Silk98, King03, Li12}.  The minimum mass required to have significant radiative feedback \citep{King03} is shown as a red dashed line in the top panel of Fig.~\ref{fig_final}. The Eddington ratio is briefly greater than one (bottom panel), but then the solid curve for the black hole growth lies below the dashed line (top panel). 
We conclude that the black hole growth is fine-tuned in such a way that the short episode of super-Eddington accretion coincides with a low black hole mass that is inadequate to curtail the accretion.

The gas infall into a halo is an essential ingredient for the formation of the galaxy as it builds a gas reservoir from which stars form \citep[e.g.][]{Lilly2013, Schaye15, Lacey2016, Vogelsberger2020,  Sharma2020, Trapp2021}. It is debatable when the infalling gas is primarily used for the growth of the black hole or when it is used for the formation of stars and the galaxy \citep[e.g.][]{DiMatteo08,  Volonteri2015b, Mo2023, Mountrichas23}. Our solution shows that, in the initial phase, the accretion rate onto the black hole is negligible; therefore, in that phase, most of the cosmological infall is likely utilised to form the galaxy. However, during the latter rapid accretion phase, the accretion rate onto the black hole is very high, and then most of the cosmological infall is likely consumed by the black hole.

The growth of the black holes in this paper should not be extrapolated for redshifts, $z\le 6$, since that would indicate unusually high black hole masses ($\ge 10^9$~M$_\odot$). It is well known that the giant black holes with Eddington ratios close to unity drive outflows, and as a result, their growth is limited \citep[e.g.][]{Churazov2005, King2010, Bischetti2022, Capelo2023}. The AGN radiation-driven outflows are a remedy to explain the mismatch between the halo mass function and the galaxy luminosity function at the high mass end \citep{Benson2003, Bower2006, Croton2006, Wechsler2018}. The quasars, including the ones recently detected by the JWST, exhibit signatures of outflows \citep{Bischetti2022, Cresci2023, Veilleux2023}. The outflows can inhibit the cosmological infall and hamper the star formation as well \citep[e.g.][]{Bower17, Costa2018}. For the highest mass SMBHs, a delicate balance likely develops between the growth of the SMBH, halo, and the galaxy \citep[e.g.][]{Sharma2020} that leads to a secular co-evolutionary behaviour \citep[e.g.][]{Gebhardt2000, Ferrarese02, Heckman2014, Li2022}.

\section{Summary and Conclusions}

We have studied a spherically symmetric accretion flow into the combined potential of a black hole at the centre of a dark matter halo. We have considered both a Hernquist and an NFW potential for the halo. The accretion flow exhibits two critical points without a central black hole. The trans-sonic flow passes through the outer critical point, then loops around the inner critical point, and does not proceed all the way to the centre.

The flow has three critical points in the presence of a black hole. The trans-sonic solution is possible through the innermost or the outermost critical point. The two solutions operate exclusively in two separate regimes of $\alpha$, roughly the ratio of the halo velocity dispersion and the sound speed. For a constant temperature, $\alpha$ decreases with redshift. Initially, due to a higher value of $\alpha$, the trans-sonic flow proceeds through the innermost critical point with a lower accretion rate. Subsequently, as $\alpha$ decreases below a transition value, the flow proceeds through the outermost critical point at a relatively high and constant dimensionless accretion rate. 

We provide the general expressions for the mass accretion rates in the combined potential of a black hole and a dark matter halo for both a Hernquist (Eq.~\ref{eq_Macc_xc}) and an NFW halo (Eq.~\ref{eq_mass_acc_nfw}). The mass accretion rate has two distinct phases, depending on whether the trans-sonic solution proceeds through the innermost critical point (due to the black hole) or the outermost critical point (due to the halo). The two phases of accretion for the SMBH are proposed in simulation studies as well, albeit for different reasons that connect SMBH growth to galaxy formation \citep{Dubois15, Angles17} and mass reservoir in the halo \citep[e.g.][]{Bower17, McAlpine18}. Moreover, the seed black hole mass used in the simulations is much higher ($\sim 10^5$~M$_\odot$) than the stellar mass seed considered in this paper. We emphasize that our finding of the two phases of accretion is based on hydrodynamic grounds and of a fundamental nature. 

In the initial phase, the mass accretion rate is due to the black hole, and it is low. After the initial phase, the mass accretion rate increases sharply (at redshift $\approx 12.5$ for a halo of present-day mass $10^{13}$~M$_\odot$). The growth of black holes to supermassive scales occurs during this phase (Fig.~\ref{fig_final} top panel).

We also calculate the evolution of the Eddington ratio and find that the accretion is mostly sub-Eddington except for a short `spiky' super-Eddington episode that occurs roughly when the mass accretion transitions from a low to a high rate (Fig.~\ref{fig_final} bottom panel). Such episodes are known to alleviate the issues in explaining high redshift SMBHs. This feature was speculated in previous studies \citep[e.g.][]{Li12}. In this paper, we have provided a firm theoretical basis for the existence of such an episode and the subsequent evolution of the Eddington ratio.

Although the super-Eddington episode is brief, it may hinder accretion through increased radiative feedback. However, during the short super-Eddington period, the black hole mass is low and likely insufficient (Fig.~\ref{fig_final} top panel) to curtail accretion through radiation feedback and outflows \citep{Silk98, King03, Li12}.     

Our solution is for an isothermal case, whereas the exact thermodynamics of the flow is likely more involved, with cooling and shock heating, which require detailed investigation. Nevertheless, our results highlight a fundamental characteristic of hydrodynamic accretion that is likely to re-appear in a general polytropic case or in a case with cooling and heating.   

The actual accretion in the haloes is likely filamentary, as often reported in cosmological simulations, with further virialization and gas cooling effects. Also, a part of the cosmological infall leads to the development of the galaxy and its spheroidal. 
Nonetheless, the mass accretion rate we provide (Eq.~\ref{eq_lamb_nfw}) is relevant for the sub-grid modelling of accretion and growth of SMBHs in cosmological simulations, as has been the standard result of Bondi \citep{Bondi1952}. The evolution of the black hole and the in-situ development of galactic components need investigation in the context of two hydrodynamic phases of accretion flow reported in this study. 

 We conclude that the stellar mass seed black holes in haloes grow via two phases of trans-sonic accretion, beginning with a phase of a low accretion rate and negligible growth due to the trans-sonic flow through the innermost critical point, followed by a phase of a high accretion rate due to the trans-sonic flow through the outermost critical point. The accretion is briefly super-Eddington, for roughly $100$~Myr, during the transition between the two phases. During the short super-Eddington episode, the accretion is unlikely to stop as the black hole mass is too low to have significant radiative feedback. Subsequently, in the phase of rapid growth, the accretion becomes sub-Eddington as the black hole grows to supermassive scales.  
 The growth history of black holes reported in this paper explains the origin of the SMBHs detected at $z \ge 6$ by the JWST  \citep{Larson23, Maiolino1_23, Maiolino2_23, Kokorev23} and quasars observed at $z\ge 6$ \citep{Fan01, Fan03, Mortlock11, Banados16, Banados18, Reed19, matsuoka18a, Matsuoka18b, Matsuoka19b, Yang20, Wang21, Inayoshi20}. 
\vspace{-0.5cm}
\section*{Acknowledgements}
We thank an anonymous referee for constructive comments that improved the manuscript. MS thanks the Department of Science and Technology (DST) and Science and Engineering Research Board (SERB) India for supporting the research through grants SRG/2022/001137 and MTR/2022/000548. RS is grateful to the Ministry of Education (MoE) and DST India for supporting this work. 

\vspace{-0.5cm}
\section*{Data Availability}
The calculations and data underlying this article will be shared upon reasonable request to the corresponding author.



\bibliography{refs} 

\begin{thebibliography}{}
\makeatletter
\relax
\def\mn@urlcharsother{\let\do\@makeother \do\$\do\&\do\#\do\^\do\_\do\%\do\~}
\def\mn@doi{\begingroup\mn@urlcharsother \@ifnextchar [ {\mn@doi@}
  {\mn@doi@[]}}
\def\mn@doi@[#1]#2{\def\@tempa{#1}\ifx\@tempa\@empty \href
  {http://dx.doi.org/#2} {doi:#2}\else \href {http://dx.doi.org/#2} {#1}\fi
  \endgroup}
\def\mn@eprint#1#2{\mn@eprint@#1:#2::\@nil}
\def\mn@eprint@arXiv#1{\href {http://arxiv.org/abs/#1} {{\tt arXiv:#1}}}
\def\mn@eprint@dblp#1{\href {http://dblp.uni-trier.de/rec/bibtex/#1.xml}
  {dblp:#1}}
\def\mn@eprint@#1:#2:#3:#4\@nil{\def\@tempa {#1}\def\@tempb {#2}\def\@tempc
  {#3}\ifx \@tempc \@empty \let \@tempc \@tempb \let \@tempb \@tempa \fi \ifx
  \@tempb \@empty \def\@tempb {arXiv}\fi \@ifundefined
  {mn@eprint@\@tempb}{\@tempb:\@tempc}{\expandafter \expandafter \csname
  mn@eprint@\@tempb\endcsname \expandafter{\@tempc}}}

\bibitem[\protect\citeauthoryear{{Abramowicz} \& {Fragile}}{{Abramowicz} \&
  {Fragile}}{2013}]{Abram13}
{Abramowicz} M.~A.,  {Fragile} P.~C.,  2013, \mn@doi [Living Reviews in
  Relativity] {10.12942/lrr-2013-1}, \href
  {https://ui.adsabs.harvard.edu/abs/2013LRR....16....1A} {16, 1}

\bibitem[\protect\citeauthoryear{{Angl{\'e}s-Alc{\'a}zar},
  {Faucher-Gigu{\`e}re}, {Quataert}, {Hopkins}, {Feldmann}, {Torrey}, {Wetzel}
  \& {Kere{\v{s}}}}{{Angl{\'e}s-Alc{\'a}zar} et~al.}{2017}]{Angles17}
{Angl{\'e}s-Alc{\'a}zar} D.,  {Faucher-Gigu{\`e}re} C.-A.,  {Quataert} E.,
  {Hopkins} P.~F.,  {Feldmann} R.,  {Torrey} P.,  {Wetzel} A.,   {Kere{\v{s}}}
  D.,  2017, \mn@doi [\mnras] {10.1093/mnrasl/slx161}, \href
  {https://ui.adsabs.harvard.edu/abs/2017MNRAS.472L.109A} {472, L109}

\bibitem[\protect\citeauthoryear{{Ba{\~n}ados} et~al.,}{{Ba{\~n}ados}
  et~al.}{2016}]{Banados16}
{Ba{\~n}ados} E.,  et~al., 2016, \mn@doi [\apjs] {10.3847/0067-0049/227/1/11},
  \href {https://ui.adsabs.harvard.edu/abs/2016ApJS..227...11B} {227, 11}

\bibitem[\protect\citeauthoryear{{Ba{\~n}ados} et~al.,}{{Ba{\~n}ados}
  et~al.}{2018}]{Banados18}
{Ba{\~n}ados} E.,  et~al., 2018, \mn@doi [\nat] {10.1038/nature25180}, \href
  {https://ui.adsabs.harvard.edu/abs/2018Natur.553..473B} {553, 473}

\bibitem[\protect\citeauthoryear{{Begelman}}{{Begelman}}{1979}]{Begelman79}
{Begelman} M.~C.,  1979, \mn@doi [\mnras] {10.1093/mnras/187.2.237}, \href
  {https://ui.adsabs.harvard.edu/abs/1979MNRAS.187..237B} {187, 237}

\bibitem[\protect\citeauthoryear{{Benson}, {Bower}, {Frenk}, {Lacey}, {Baugh}
  \& {Cole}}{{Benson} et~al.}{2003}]{Benson2003}
{Benson} A.~J.,  {Bower} R.~G.,  {Frenk} C.~S.,  {Lacey} C.~G.,  {Baugh} C.~M.,
    {Cole} S.,  2003, \mn@doi [\apj] {10.1086/379160}, \href
  {https://ui.adsabs.harvard.edu/abs/2003ApJ...599...38B} {599, 38}

\bibitem[\protect\citeauthoryear{{Birnboim} \& {Dekel}}{{Birnboim} \&
  {Dekel}}{2003}]{Birnboim03}
{Birnboim} Y.,  {Dekel} A.,  2003, \mn@doi [\mnras]
  {10.1046/j.1365-8711.2003.06955.x}, \href
  {https://ui.adsabs.harvard.edu/abs/2003MNRAS.345..349B} {345, 349}

\bibitem[\protect\citeauthoryear{{Bischetti} et~al.,}{{Bischetti}
  et~al.}{2022}]{Bischetti2022}
{Bischetti} M.,  et~al., 2022, \mn@doi [\nat] {10.1038/s41586-022-04608-1},
  \href {https://ui.adsabs.harvard.edu/abs/2022Natur.605..244B} {605, 244}

\bibitem[\protect\citeauthoryear{{Bondi}}{{Bondi}}{1952}]{Bondi1952}
{Bondi} H.,  1952, \mn@doi [\mnras] {10.1093/mnras/112.2.195}, \href
  {https://ui.adsabs.harvard.edu/abs/1952MNRAS.112..195B} {112, 195}

\bibitem[\protect\citeauthoryear{{Booth} \& {Schaye}}{{Booth} \&
  {Schaye}}{2009}]{Booth2009}
{Booth} C.~M.,  {Schaye} J.,  2009, \mn@doi [\mnras]
  {10.1111/j.1365-2966.2009.15043.x}, \href
  {https://ui.adsabs.harvard.edu/abs/2009MNRAS.398...53B} {398, 53}

\bibitem[\protect\citeauthoryear{{Booth} \& {Schaye}}{{Booth} \&
  {Schaye}}{2010}]{Booth10}
{Booth} C.~M.,  {Schaye} J.,  2010, \mn@doi [\mnras]
  {10.1111/j.1745-3933.2010.00832.x}, \href
  {https://ui.adsabs.harvard.edu/abs/2010MNRAS.405L...1B} {405, L1}

\bibitem[\protect\citeauthoryear{{Bower}, {Benson}, {Malbon}, {Helly}, {Frenk},
  {Baugh}, {Cole}  \& {Lacey}}{{Bower} et~al.}{2006}]{Bower2006}
{Bower} R.~G.,  {Benson} A.~J.,  {Malbon} R.,  {Helly} J.~C.,  {Frenk} C.~S.,
  {Baugh} C.~M.,  {Cole} S.,   {Lacey} C.~G.,  2006, \mn@doi [\mnras]
  {10.1111/j.1365-2966.2006.10519.x}, \href
  {https://ui.adsabs.harvard.edu/abs/2006MNRAS.370..645B} {370, 645}

\bibitem[\protect\citeauthoryear{{Bower}, {Schaye}, {Frenk}, {Theuns},
  {Schaller}, {Crain}  \& {McAlpine}}{{Bower} et~al.}{2017}]{Bower17}
{Bower} R.~G.,  {Schaye} J.,  {Frenk} C.~S.,  {Theuns} T.,  {Schaller} M.,
  {Crain} R.~A.,   {McAlpine} S.,  2017, \mn@doi [\mnras]
  {10.1093/mnras/stw2735}, \href
  {https://ui.adsabs.harvard.edu/abs/2017MNRAS.465...32B} {465, 32}

\bibitem[\protect\citeauthoryear{{Bromm} \& {Loeb}}{{Bromm} \&
  {Loeb}}{2003}]{Bromm03}
{Bromm} V.,  {Loeb} A.,  2003, \mn@doi [\apj] {10.1086/377529}, \href
  {https://ui.adsabs.harvard.edu/abs/2003ApJ...596...34B} {596, 34}

\bibitem[\protect\citeauthoryear{{Capelo}, {Feruglio}, {Hickox}  \&
  {Tombesi}}{{Capelo} et~al.}{2023}]{Capelo2023}
{Capelo} P.~R.,  {Feruglio} C.,  {Hickox} R.~C.,   {Tombesi} F.,  2023, in ,
  Handbook of X-ray and Gamma-ray Astrophysics. Edited by Cosimo Bambi and
  Andrea Santangelo.
p.~126, \mn@doi{10.1007/978-981-16-4544-0_115-1}

\bibitem[\protect\citeauthoryear{{Churazov}, {Sazonov}, {Sunyaev}, {Forman},
  {Jones}  \& {B{\"o}hringer}}{{Churazov} et~al.}{2005}]{Churazov2005}
{Churazov} E.,  {Sazonov} S.,  {Sunyaev} R.,  {Forman} W.,  {Jones} C.,
  {B{\"o}hringer} H.,  2005, \mn@doi [\mnras]
  {10.1111/j.1745-3933.2005.00093.x}, \href
  {https://ui.adsabs.harvard.edu/abs/2005MNRAS.363L..91C} {363, L91}

\bibitem[\protect\citeauthoryear{{Ciotti} \& {Pellegrini}}{{Ciotti} \&
  {Pellegrini}}{2017}]{Ciotti17}
{Ciotti} L.,  {Pellegrini} S.,  2017, \mn@doi [\apj]
  {10.3847/1538-4357/aa8d1f}, \href
  {https://ui.adsabs.harvard.edu/abs/2017ApJ...848...29C} {848, 29}

\bibitem[\protect\citeauthoryear{{Ciotti} \& {Pellegrini}}{{Ciotti} \&
  {Pellegrini}}{2018}]{Ciotti18}
{Ciotti} L.,  {Pellegrini} S.,  2018, \mn@doi [\apj]
  {10.3847/1538-4357/aae97d}, \href
  {https://ui.adsabs.harvard.edu/abs/2018ApJ...868...91C} {868, 91}

\bibitem[\protect\citeauthoryear{{Collin} \& {Kawaguchi}}{{Collin} \&
  {Kawaguchi}}{2004}]{Collin04}
{Collin} S.,  {Kawaguchi} T.,  2004, \mn@doi [\aap]
  {10.1051/0004-6361:20040528}, \href
  {https://ui.adsabs.harvard.edu/abs/2004A&A...426..797C} {426, 797}

\bibitem[\protect\citeauthoryear{{Correa}, {Wyithe}, {Schaye}  \&
  {Duffy}}{{Correa} et~al.}{2015a}]{Correa1}
{Correa} C.~A.,  {Wyithe} J. S.~B.,  {Schaye} J.,   {Duffy} A.~R.,  2015a,
  \mn@doi [\mnras] {10.1093/mnras/stv689}, \href
  {https://ui.adsabs.harvard.edu/abs/2015MNRAS.450.1514C} {450, 1514}

\bibitem[\protect\citeauthoryear{{Correa}, {Wyithe}, {Schaye}  \&
  {Duffy}}{{Correa} et~al.}{2015b}]{Correa3}
{Correa} C.~A.,  {Wyithe} J. S.~B.,  {Schaye} J.,   {Duffy} A.~R.,  2015b,
  \mn@doi [\mnras] {10.1093/mnras/stv1363}, \href
  {https://ui.adsabs.harvard.edu/abs/2015MNRAS.452.1217C} {452, 1217}

\bibitem[\protect\citeauthoryear{{Costa}, {Rosdahl}, {Sijacki}  \&
  {Haehnelt}}{{Costa} et~al.}{2018}]{Costa2018}
{Costa} T.,  {Rosdahl} J.,  {Sijacki} D.,   {Haehnelt} M.~G.,  2018, \mn@doi
  [\mnras] {10.1093/mnras/sty1514}, \href
  {https://ui.adsabs.harvard.edu/abs/2018MNRAS.479.2079C} {479, 2079}

\bibitem[\protect\citeauthoryear{{Cresci} et~al.,}{{Cresci}
  et~al.}{2023}]{Cresci2023}
{Cresci} G.,  et~al., 2023, \mn@doi [\aap] {10.1051/0004-6361/202346001}, \href
  {https://ui.adsabs.harvard.edu/abs/2023A&A...672A.128C} {672, A128}

\bibitem[\protect\citeauthoryear{{Croft}, {Di Matteo}, {Springel}  \&
  {Hernquist}}{{Croft} et~al.}{2009}]{Croft09}
{Croft} R. A.~C.,  {Di Matteo} T.,  {Springel} V.,   {Hernquist} L.,  2009,
  \mn@doi [\mnras] {10.1111/j.1365-2966.2009.15446.x}, \href
  {https://ui.adsabs.harvard.edu/abs/2009MNRAS.400...43C} {400, 43}

\bibitem[\protect\citeauthoryear{{Croton} et~al.,}{{Croton}
  et~al.}{2006}]{Croton2006}
{Croton} D.~J.,  et~al., 2006, \mn@doi [\mnras]
  {10.1111/j.1365-2966.2005.09675.x}, \href
  {https://ui.adsabs.harvard.edu/abs/2006MNRAS.365...11C} {365, 11}

\bibitem[\protect\citeauthoryear{{Dekel} et~al.,}{{Dekel}
  et~al.}{2009}]{Dekel09}
{Dekel} A.,  et~al., 2009, \mn@doi [\nat] {10.1038/nature07648}, \href
  {https://ui.adsabs.harvard.edu/abs/2009Natur.457..451D} {457, 451}

\bibitem[\protect\citeauthoryear{{Di Matteo}, {Colberg}, {Springel},
  {Hernquist}  \& {Sijacki}}{{Di Matteo} et~al.}{2008}]{DiMatteo08}
{Di Matteo} T.,  {Colberg} J.,  {Springel} V.,  {Hernquist} L.,   {Sijacki} D.,
   2008, \mn@doi [\apj] {10.1086/524921}, \href
  {https://ui.adsabs.harvard.edu/abs/2008ApJ...676...33D} {676, 33}

\bibitem[\protect\citeauthoryear{{Du} et~al.,}{{Du} et~al.}{2014}]{Du14}
{Du} P.,  et~al., 2014, \mn@doi [\apj] {10.1088/0004-637X/782/1/45}, \href
  {https://ui.adsabs.harvard.edu/abs/2014ApJ...782...45D} {782, 45}

\bibitem[\protect\citeauthoryear{{Dubois}, {Volonteri}, {Silk}, {Devriendt},
  {Slyz}  \& {Teyssier}}{{Dubois} et~al.}{2015}]{Dubois15}
{Dubois} Y.,  {Volonteri} M.,  {Silk} J.,  {Devriendt} J.,  {Slyz} A.,
  {Teyssier} R.,  2015, \mn@doi [\mnras] {10.1093/mnras/stv1416}, \href
  {https://ui.adsabs.harvard.edu/abs/2015MNRAS.452.1502D} {452, 1502}

\bibitem[\protect\citeauthoryear{{Fan} et~al.,}{{Fan} et~al.}{2001}]{Fan01}
{Fan} X.,  et~al., 2001, \mn@doi [\aj] {10.1086/324111}, \href
  {https://ui.adsabs.harvard.edu/abs/2001AJ....122.2833F} {122, 2833}

\bibitem[\protect\citeauthoryear{{Fan} et~al.,}{{Fan} et~al.}{2003}]{Fan03}
{Fan} X.,  et~al., 2003, \mn@doi [\aj] {10.1086/368246}, \href
  {https://ui.adsabs.harvard.edu/abs/2003AJ....125.1649F} {125, 1649}

\bibitem[\protect\citeauthoryear{{Ferrarese}}{{Ferrarese}}{2002}]{Ferrarese02}
{Ferrarese} L.,  2002, \mn@doi [\apj] {10.1086/342308}, \href
  {https://ui.adsabs.harvard.edu/abs/2002ApJ...578...90F} {578, 90}

\bibitem[\protect\citeauthoryear{{Gebhardt} et~al.,}{{Gebhardt}
  et~al.}{2000}]{Gebhardt2000}
{Gebhardt} K.,  et~al., 2000, \mn@doi [\apjl] {10.1086/312840}, \href
  {https://ui.adsabs.harvard.edu/abs/2000ApJ...539L..13G} {539, L13}

\bibitem[\protect\citeauthoryear{{Habouzit}, {Volonteri}  \&
  {Dubois}}{{Habouzit} et~al.}{2017}]{Habouzit17}
{Habouzit} M.,  {Volonteri} M.,   {Dubois} Y.,  2017, \mn@doi [\mnras]
  {10.1093/mnras/stx666}, \href
  {https://ui.adsabs.harvard.edu/abs/2017MNRAS.468.3935H} {468, 3935}

\bibitem[\protect\citeauthoryear{{Haiman} \& {Loeb}}{{Haiman} \&
  {Loeb}}{2001}]{Haiman01}
{Haiman} Z.,  {Loeb} A.,  2001, \mn@doi [\apj] {10.1086/320586}, \href
  {https://ui.adsabs.harvard.edu/abs/2001ApJ...552..459H} {552, 459}

\bibitem[\protect\citeauthoryear{{H{\"a}ring} \& {Rix}}{{H{\"a}ring} \&
  {Rix}}{2004}]{Haring04}
{H{\"a}ring} N.,  {Rix} H.-W.,  2004, \mn@doi [\apjl] {10.1086/383567}, \href
  {https://ui.adsabs.harvard.edu/abs/2004ApJ...604L..89H} {604, L89}

\bibitem[\protect\citeauthoryear{{Heckman} \& {Best}}{{Heckman} \&
  {Best}}{2014}]{Heckman2014}
{Heckman} T.~M.,  {Best} P.~N.,  2014, \mn@doi [\araa]
  {10.1146/annurev-astro-081913-035722}, \href
  {https://ui.adsabs.harvard.edu/abs/2014ARA&A..52..589H} {52, 589}

\bibitem[\protect\citeauthoryear{{Hernquist}}{{Hernquist}}{1990}]{Hernquist90}
{Hernquist} L.,  1990, \mn@doi [\apj] {10.1086/168845}, \href
  {https://ui.adsabs.harvard.edu/abs/1990ApJ...356..359H} {356, 359}

\bibitem[\protect\citeauthoryear{Hobbs, Power, Nayakshin  \& King}{Hobbs
  et~al.}{2012}]{Hobbs2012}
Hobbs A.,  Power C.,  Nayakshin S.,   King A.~R.,  2012, Monthly Notices of the
  Royal Astronomical Society, 421, 3443

\bibitem[\protect\citeauthoryear{{Hopkins}, {Hernquist}, {Cox}  \&
  {Kere{\v{s}}}}{{Hopkins} et~al.}{2008}]{Hopkins08}
{Hopkins} P.~F.,  {Hernquist} L.,  {Cox} T.~J.,   {Kere{\v{s}}} D.,  2008,
  \mn@doi [\apjs] {10.1086/524362}, \href
  {https://ui.adsabs.harvard.edu/abs/2008ApJS..175..356H} {175, 356}

\bibitem[\protect\citeauthoryear{{Inayoshi}, {Visbal}  \& {Haiman}}{{Inayoshi}
  et~al.}{2020}]{Inayoshi20}
{Inayoshi} K.,  {Visbal} E.,   {Haiman} Z.,  2020, \mn@doi [\araa]
  {10.1146/annurev-astro-120419-014455}, \href
  {https://ui.adsabs.harvard.edu/abs/2020ARA&A..58...27I} {58, 27}

\bibitem[\protect\citeauthoryear{{Jaffe}}{{Jaffe}}{1983}]{Jaffe1983}
{Jaffe} W.,  1983, \mn@doi [\mnras] {10.1093/mnras/202.4.995}, \href
  {https://ui.adsabs.harvard.edu/abs/1983MNRAS.202..995J} {202, 995}

\bibitem[\protect\citeauthoryear{{Johnson} \& {Bromm}}{{Johnson} \&
  {Bromm}}{2007}]{Johnson07}
{Johnson} J.~L.,  {Bromm} V.,  2007, \mn@doi [\mnras]
  {10.1111/j.1365-2966.2006.11275.x}, \href
  {https://ui.adsabs.harvard.edu/abs/2007MNRAS.374.1557J} {374, 1557}

\bibitem[\protect\citeauthoryear{{Kawasaki}, {Kusenko}  \&
  {Yanagida}}{{Kawasaki} et~al.}{2012}]{Kawasaki12}
{Kawasaki} M.,  {Kusenko} A.,   {Yanagida} T.~T.,  2012, \mn@doi [Physics
  Letters B] {10.1016/j.physletb.2012.03.056}, \href
  {https://ui.adsabs.harvard.edu/abs/2012PhLB..711....1K} {711, 1}

\bibitem[\protect\citeauthoryear{{King}}{{King}}{2003}]{King03}
{King} A.,  2003, \mn@doi [\apjl] {10.1086/379143}, \href
  {https://ui.adsabs.harvard.edu/abs/2003ApJ...596L..27K} {596, L27}

\bibitem[\protect\citeauthoryear{{King}}{{King}}{2010}]{King2010}
{King} A.~R.,  2010, \mn@doi [\mnras] {10.1111/j.1365-2966.2009.16013.x}, \href
  {https://ui.adsabs.harvard.edu/abs/2010MNRAS.402.1516K} {402, 1516}

\bibitem[\protect\citeauthoryear{{Kokorev} et~al.,}{{Kokorev}
  et~al.}{2023}]{Kokorev23}
{Kokorev} V.,  et~al., 2023, \mn@doi [\apjl] {10.3847/2041-8213/ad037a}, \href
  {https://ui.adsabs.harvard.edu/abs/2023ApJ...957L...7K} {957, L7}

\bibitem[\protect\citeauthoryear{{Kormendy} \& {Ho}}{{Kormendy} \&
  {Ho}}{2013}]{Kormendy13}
{Kormendy} J.,  {Ho} L.~C.,  2013, \mn@doi [\araa]
  {10.1146/annurev-astro-082708-101811}, \href
  {https://ui.adsabs.harvard.edu/abs/2013ARA&A..51..511K} {51, 511}

\bibitem[\protect\citeauthoryear{{Kormendy} \& {Richstone}}{{Kormendy} \&
  {Richstone}}{1995}]{Kormendy95}
{Kormendy} J.,  {Richstone} D.,  1995, \mn@doi [\araa]
  {10.1146/annurev.aa.33.090195.003053}, \href
  {https://ui.adsabs.harvard.edu/abs/1995ARA&A..33..581K} {33, 581}

\bibitem[\protect\citeauthoryear{{Lacey} et~al.,}{{Lacey}
  et~al.}{2016}]{Lacey2016}
{Lacey} C.~G.,  et~al., 2016, \mn@doi [\mnras] {10.1093/mnras/stw1888}, \href
  {https://ui.adsabs.harvard.edu/abs/2016MNRAS.462.3854L} {462, 3854}

\bibitem[\protect\citeauthoryear{{Larson} et~al.,}{{Larson}
  et~al.}{2023}]{Larson23}
{Larson} R.~L.,  et~al., 2023, \mn@doi [\apjl] {10.3847/2041-8213/ace619},
  \href {https://ui.adsabs.harvard.edu/abs/2023ApJ...953L..29L} {953, L29}

\bibitem[\protect\citeauthoryear{{Li}}{{Li}}{2012}]{Li12}
{Li} L.-X.,  2012, \mn@doi [\mnras] {10.1111/j.1365-2966.2012.21336.x}, \href
  {https://ui.adsabs.harvard.edu/abs/2012MNRAS.424.1461L} {424, 1461}

\bibitem[\protect\citeauthoryear{{Li} et~al.,}{{Li} et~al.}{2022}]{Li2022}
{Li} J.,  et~al., 2022, \mn@doi [\apjl] {10.3847/2041-8213/ac6de8}, \href
  {https://ui.adsabs.harvard.edu/abs/2022ApJ...931L..11L} {931, L11}

\bibitem[\protect\citeauthoryear{{Lilly}, {Carollo}, {Pipino}, {Renzini}  \&
  {Peng}}{{Lilly} et~al.}{2013}]{Lilly2013}
{Lilly} S.~J.,  {Carollo} C.~M.,  {Pipino} A.,  {Renzini} A.,   {Peng} Y.,
  2013, \mn@doi [\apj] {10.1088/0004-637X/772/2/119}, \href
  {https://ui.adsabs.harvard.edu/abs/2013ApJ...772..119L} {772, 119}

\bibitem[\protect\citeauthoryear{{Lynden-Bell}}{{Lynden-Bell}}{1969}]{Lynden69}
{Lynden-Bell} D.,  1969, \mn@doi [\nat] {10.1038/223690a0}, \href
  {https://ui.adsabs.harvard.edu/abs/1969Natur.223..690L} {223, 690}

\bibitem[\protect\citeauthoryear{{Magorrian} et~al.,}{{Magorrian}
  et~al.}{1998}]{Magorrian98}
{Magorrian} J.,  et~al., 1998, \mn@doi [\aj] {10.1086/300353}, \href
  {https://ui.adsabs.harvard.edu/abs/1998AJ....115.2285M} {115, 2285}

\bibitem[\protect\citeauthoryear{{Maiolino} et~al.,}{{Maiolino}
  et~al.}{2023a}]{Maiolino1_23}
{Maiolino} R.,  et~al., 2023a, \mn@doi [arXiv e-prints]
  {10.48550/arXiv.2305.12492}, \href
  {https://ui.adsabs.harvard.edu/abs/2023arXiv230512492M} {p. arXiv:2305.12492}

\bibitem[\protect\citeauthoryear{{Maiolino} et~al.,}{{Maiolino}
  et~al.}{2023b}]{Maiolino2_23}
{Maiolino} R.,  et~al., 2023b, \mn@doi [arXiv e-prints]
  {10.48550/arXiv.2308.01230}, \href
  {https://ui.adsabs.harvard.edu/abs/2023arXiv230801230M} {p. arXiv:2308.01230}

\bibitem[\protect\citeauthoryear{{Matsuoka} et~al.,}{{Matsuoka}
  et~al.}{2018a}]{Matsuoka18b}
{Matsuoka} Y.,  et~al., 2018a, \mn@doi [\pasj] {10.1093/pasj/psx046}, \href
  {https://ui.adsabs.harvard.edu/abs/2018PASJ...70S..35M} {70, S35}

\bibitem[\protect\citeauthoryear{{Matsuoka} et~al.,}{{Matsuoka}
  et~al.}{2018b}]{matsuoka18a}
{Matsuoka} Y.,  et~al., 2018b, \mn@doi [\apjs] {10.3847/1538-4365/aac724},
  \href {https://ui.adsabs.harvard.edu/abs/2018ApJS..237....5M} {237, 5}

\bibitem[\protect\citeauthoryear{{Matsuoka} et~al.,}{{Matsuoka}
  et~al.}{2019a}]{Matsuoka19b}
{Matsuoka} Y.,  et~al., 2019a, \mn@doi [\apjl] {10.3847/2041-8213/ab0216},
  \href {https://ui.adsabs.harvard.edu/abs/2019ApJ...872L...2M} {872, L2}

\bibitem[\protect\citeauthoryear{{Matsuoka} et~al.,}{{Matsuoka}
  et~al.}{2019b}]{Matsuoka19a}
{Matsuoka} Y.,  et~al., 2019b, \mn@doi [\apj] {10.3847/1538-4357/ab3c60}, \href
  {https://ui.adsabs.harvard.edu/abs/2019ApJ...883..183M} {883, 183}

\bibitem[\protect\citeauthoryear{{McAlpine}, {Bower}, {Harrison}, {Crain},
  {Schaller}, {Schaye}  \& {Theuns}}{{McAlpine} et~al.}{2017}]{McAlpine17}
{McAlpine} S.,  {Bower} R.~G.,  {Harrison} C.~M.,  {Crain} R.~A.,  {Schaller}
  M.,  {Schaye} J.,   {Theuns} T.,  2017, \mn@doi [\mnras]
  {10.1093/mnras/stx658}, \href
  {https://ui.adsabs.harvard.edu/abs/2017MNRAS.468.3395M} {468, 3395}

\bibitem[\protect\citeauthoryear{{McAlpine}, {Bower}, {Rosario}, {Crain},
  {Schaye}  \& {Theuns}}{{McAlpine} et~al.}{2018}]{McAlpine18}
{McAlpine} S.,  {Bower} R.~G.,  {Rosario} D.~J.,  {Crain} R.~A.,  {Schaye} J.,
   {Theuns} T.,  2018, \mn@doi [\mnras] {10.1093/mnras/sty2489}, \href
  {https://ui.adsabs.harvard.edu/abs/2018MNRAS.481.3118M} {481, 3118}

\bibitem[\protect\citeauthoryear{{McKinney}, {Dai}  \& {Avara}}{{McKinney}
  et~al.}{2015}]{McKinney15}
{McKinney} J.~C.,  {Dai} L.,   {Avara} M.~J.,  2015, \mn@doi [\mnras]
  {10.1093/mnrasl/slv115}, \href
  {https://ui.adsabs.harvard.edu/abs/2015MNRAS.454L...6M} {454, L6}

\bibitem[\protect\citeauthoryear{{McLure} \& {Dunlop}}{{McLure} \&
  {Dunlop}}{2002}]{Maclure02}
{McLure} R.~J.,  {Dunlop} J.~S.,  2002, \mn@doi [\mnras]
  {10.1046/j.1365-8711.2002.05236.x}, \href
  {https://ui.adsabs.harvard.edu/abs/2002MNRAS.331..795M} {331, 795}

\bibitem[\protect\citeauthoryear{{Mineshige}, {Kawaguchi}, {Takeuchi}  \&
  {Hayashida}}{{Mineshige} et~al.}{2000}]{Mineshige00}
{Mineshige} S.,  {Kawaguchi} T.,  {Takeuchi} M.,   {Hayashida} K.,  2000,
  \mn@doi [\pasj] {10.1093/pasj/52.3.499}, \href
  {https://ui.adsabs.harvard.edu/abs/2000PASJ...52..499M} {52, 499}

\bibitem[\protect\citeauthoryear{{Mo}, {Mao}  \& {White}}{{Mo}
  et~al.}{1998}]{Mo1998}
{Mo} H.~J.,  {Mao} S.,   {White} S. D.~M.,  1998, \mn@doi [\mnras]
  {10.1046/j.1365-8711.1998.01227.x}, \href
  {https://ui.adsabs.harvard.edu/abs/1998MNRAS.295..319M} {295, 319}

\bibitem[\protect\citeauthoryear{{Mo}, {Chen}  \& {Wang}}{{Mo}
  et~al.}{2023}]{Mo2023}
{Mo} H.,  {Chen} Y.,   {Wang} H.,  2023, \mn@doi [arXiv e-prints]
  {10.48550/arXiv.2311.05030}, \href
  {https://ui.adsabs.harvard.edu/abs/2023arXiv231105030M} {p. arXiv:2311.05030}

\bibitem[\protect\citeauthoryear{{Mortlock} et~al.,}{{Mortlock}
  et~al.}{2011}]{Mortlock11}
{Mortlock} D.~J.,  et~al., 2011, \mn@doi [\nat] {10.1038/nature10159}, \href
  {https://ui.adsabs.harvard.edu/abs/2011Natur.474..616M} {474, 616}

\bibitem[\protect\citeauthoryear{{Mountrichas}}{{Mountrichas}}{2023}]{Mountrichas23}
{Mountrichas} G.,  2023, \mn@doi [\aap] {10.1051/0004-6361/202345924}, \href
  {https://ui.adsabs.harvard.edu/abs/2023A&A...672A..98M} {672, A98}

\bibitem[\protect\citeauthoryear{{Narayan}, {Mahadevan}  \&
  {Quataert}}{{Narayan} et~al.}{1998}]{Narayan1998}
{Narayan} R.,  {Mahadevan} R.,   {Quataert} E.,  1998, in {Abramowicz} M.~A.,
  {Bj{\"o}rnsson} G.,   {Pringle} J.~E.,  eds, Theory of Black Hole Accretion
  Disks. pp 148--182 (\mn@eprint {arXiv} {astro-ph/9803141})

\bibitem[\protect\citeauthoryear{{Narayan}, {Igumenshchev}  \&
  {Abramowicz}}{{Narayan} et~al.}{2003}]{Narayan03}
{Narayan} R.,  {Igumenshchev} I.~V.,   {Abramowicz} M.~A.,  2003, \mn@doi
  [\pasj] {10.1093/pasj/55.6.L69}, \href
  {https://ui.adsabs.harvard.edu/abs/2003PASJ...55L..69N} {55, L69}

\bibitem[\protect\citeauthoryear{{Navarro}, {Frenk}  \& {White}}{{Navarro}
  et~al.}{1997}]{Navarro97}
{Navarro} J.~F.,  {Frenk} C.~S.,   {White} S. D.~M.,  1997, \mn@doi [\apj]
  {10.1086/304888}, \href
  {https://ui.adsabs.harvard.edu/abs/1997ApJ...490..493N} {490, 493}

\bibitem[\protect\citeauthoryear{{Okamoto}, {Nemmen}  \& {Bower}}{{Okamoto}
  et~al.}{2008}]{Okamoto08}
{Okamoto} T.,  {Nemmen} R.~S.,   {Bower} R.~G.,  2008, \mn@doi [\mnras]
  {10.1111/j.1365-2966.2008.12883.x}, \href
  {https://ui.adsabs.harvard.edu/abs/2008MNRAS.385..161O} {385, 161}

\bibitem[\protect\citeauthoryear{{Okuda}}{{Okuda}}{2002}]{Okuda02}
{Okuda} T.,  2002, \mn@doi [\pasj] {10.1093/pasj/54.2.253}, \href
  {https://ui.adsabs.harvard.edu/abs/2002PASJ...54..253O} {54, 253}

\bibitem[\protect\citeauthoryear{{Park}}{{Park}}{2017}]{Park2017}
{Park} M.~G.,  2017, in 35th International Cosmic Ray Conference (ICRC2017).
  p.~1086, \mn@doi{10.22323/1.301.01086}

\bibitem[\protect\citeauthoryear{{Park} \& {Ricotti}}{{Park} \&
  {Ricotti}}{2011}]{Park11}
{Park} K.,  {Ricotti} M.,  2011, \mn@doi [\apj] {10.1088/0004-637X/739/1/2},
  \href {https://ui.adsabs.harvard.edu/abs/2011ApJ...739....2P} {739, 2}

\bibitem[\protect\citeauthoryear{{Planck Collaboration} et~al.,}{{Planck
  Collaboration} et~al.}{2020}]{Planck20}
{Planck Collaboration} et~al., 2020, \mn@doi [\aap]
  {10.1051/0004-6361/201833910}, \href
  {https://ui.adsabs.harvard.edu/abs/2020A&A...641A...6P} {641, A6}

\bibitem[\protect\citeauthoryear{{Prada}, {Klypin}, {Cuesta}, {Betancort-Rijo}
  \& {Primack}}{{Prada} et~al.}{2012}]{Prada12}
{Prada} F.,  {Klypin} A.~A.,  {Cuesta} A.~J.,  {Betancort-Rijo} J.~E.,
  {Primack} J.,  2012, \mn@doi [\mnras] {10.1111/j.1365-2966.2012.21007.x},
  \href {https://ui.adsabs.harvard.edu/abs/2012MNRAS.423.3018P} {423, 3018}

\bibitem[\protect\citeauthoryear{{Pringle}}{{Pringle}}{1981}]{Pringle81}
{Pringle} J.~E.,  1981, \mn@doi [\araa] {10.1146/annurev.aa.19.090181.001033},
  \href {https://ui.adsabs.harvard.edu/abs/1981ARA&A..19..137P} {19, 137}

\bibitem[\protect\citeauthoryear{{Reed} et~al.,}{{Reed} et~al.}{2019}]{Reed19}
{Reed} S.~L.,  et~al., 2019, \mn@doi [\mnras] {10.1093/mnras/stz1341}, \href
  {https://ui.adsabs.harvard.edu/abs/2019MNRAS.487.1874R} {487, 1874}

\bibitem[\protect\citeauthoryear{{Rees}}{{Rees}}{1984}]{Rees84}
{Rees} M.~J.,  1984, \mn@doi [\araa] {10.1146/annurev.aa.22.090184.002351},
  \href {https://ui.adsabs.harvard.edu/abs/1984ARA&A..22..471R} {22, 471}

\bibitem[\protect\citeauthoryear{{Regan}, {Downes}, {Volonteri}, {Beckmann},
  {Lupi}, {Trebitsch}  \& {Dubois}}{{Regan} et~al.}{2019}]{Regan19}
{Regan} J.~A.,  {Downes} T.~P.,  {Volonteri} M.,  {Beckmann} R.,  {Lupi} A.,
  {Trebitsch} M.,   {Dubois} Y.,  2019, \mn@doi [\mnras]
  {10.1093/mnras/stz1045}, \href
  {https://ui.adsabs.harvard.edu/abs/2019MNRAS.486.3892R} {486, 3892}

\bibitem[\protect\citeauthoryear{{Richstone} et~al.,}{{Richstone}
  et~al.}{1998}]{Richstone18}
{Richstone} D.,  et~al., 1998, \mn@doi [\nat]
  {10.48550/arXiv.astro-ph/9810378}, \href
  {https://ui.adsabs.harvard.edu/abs/1998Natur.395A..14R} {385, A14}

\bibitem[\protect\citeauthoryear{{Robertson}, {Hernquist}, {Cox}, {Di Matteo},
  {Hopkins}, {Martini}  \& {Springel}}{{Robertson} et~al.}{2006}]{Robertson06}
{Robertson} B.,  {Hernquist} L.,  {Cox} T.~J.,  {Di Matteo} T.,  {Hopkins}
  P.~F.,  {Martini} P.,   {Springel} V.,  2006, \mn@doi [\apj]
  {10.1086/500348}, \href
  {https://ui.adsabs.harvard.edu/abs/2006ApJ...641...90R} {641, 90}

\bibitem[\protect\citeauthoryear{{Rosas-Guevara} et~al.,}{{Rosas-Guevara}
  et~al.}{2015}]{Rosas15}
{Rosas-Guevara} Y.~M.,  et~al., 2015, \mn@doi [\mnras] {10.1093/mnras/stv2056},
  \href {https://ui.adsabs.harvard.edu/abs/2015MNRAS.454.1038R} {454, 1038}

\bibitem[\protect\citeauthoryear{{Rosas-Guevara}, {Bower}, {Schaye},
  {McAlpine}, {Dalla Vecchia}, {Frenk}, {Schaller}  \&
  {Theuns}}{{Rosas-Guevara} et~al.}{2016}]{Rosas16}
{Rosas-Guevara} Y.,  {Bower} R.~G.,  {Schaye} J.,  {McAlpine} S.,  {Dalla
  Vecchia} C.,  {Frenk} C.~S.,  {Schaller} M.,   {Theuns} T.,  2016, \mn@doi
  [\mnras] {10.1093/mnras/stw1679}, \href
  {https://ui.adsabs.harvard.edu/abs/2016MNRAS.462..190R} {462, 190}

\bibitem[\protect\citeauthoryear{{Schaye} et~al.,}{{Schaye}
  et~al.}{2015}]{Schaye15}
{Schaye} J.,  et~al., 2015, \mn@doi [\mnras] {10.1093/mnras/stu2058}, \href
  {https://ui.adsabs.harvard.edu/abs/2015MNRAS.446..521S} {446, 521}

\bibitem[\protect\citeauthoryear{{Shankar}, {Salucci}, {Granato}, {De Zotti}
  \& {Danese}}{{Shankar} et~al.}{2004}]{Shankar04}
{Shankar} F.,  {Salucci} P.,  {Granato} G.~L.,  {De Zotti} G.,   {Danese} L.,
  2004, \mn@doi [\mnras] {10.1111/j.1365-2966.2004.08261.x}, \href
  {https://ui.adsabs.harvard.edu/abs/2004MNRAS.354.1020S} {354, 1020}

\bibitem[\protect\citeauthoryear{{Sharma} \& {Theuns}}{{Sharma} \&
  {Theuns}}{2020}]{Sharma2020}
{Sharma} M.,  {Theuns} T.,  2020, \mn@doi [\mnras] {10.1093/mnras/stz2909},
  \href {https://ui.adsabs.harvard.edu/abs/2020MNRAS.492.2418S} {492, 2418}

\bibitem[\protect\citeauthoryear{{Silk} \& {Rees}}{{Silk} \&
  {Rees}}{1998}]{Silk98}
{Silk} J.,  {Rees} M.~J.,  1998, \mn@doi [\aap]
  {10.48550/arXiv.astro-ph/9801013}, \href
  {https://ui.adsabs.harvard.edu/abs/1998A&A...331L...1S} {331, L1}

\bibitem[\protect\citeauthoryear{{S{\k{a}}dowski}, {Narayan}, {Tchekhovskoy},
  {Abarca}, {Zhu}  \& {McKinney}}{{S{\k{a}}dowski} et~al.}{2015}]{Sadowski15}
{S{\k{a}}dowski} A.,  {Narayan} R.,  {Tchekhovskoy} A.,  {Abarca} D.,  {Zhu}
  Y.,   {McKinney} J.~C.,  2015, \mn@doi [\mnras] {10.1093/mnras/stu2387},
  \href {https://ui.adsabs.harvard.edu/abs/2015MNRAS.447...49S} {447, 49}

\bibitem[\protect\citeauthoryear{{Smith} \& {Bromm}}{{Smith} \&
  {Bromm}}{2019}]{Smith19}
{Smith} A.,  {Bromm} V.,  2019, \mn@doi [Contemporary Physics]
  {10.1080/00107514.2019.1615715}, \href
  {https://ui.adsabs.harvard.edu/abs/2019ConPh..60..111S} {60, 111}

\bibitem[\protect\citeauthoryear{{Soltan}}{{Soltan}}{1982}]{Soltan82}
{Soltan} A.,  1982, \mn@doi [\mnras] {10.1093/mnras/200.1.115}, \href
  {https://ui.adsabs.harvard.edu/abs/1982MNRAS.200..115S} {200, 115}

\bibitem[\protect\citeauthoryear{{Springel} et~al.,}{{Springel}
  et~al.}{2005}]{Springel05}
{Springel} V.,  et~al., 2005, \mn@doi [\nat] {10.1038/nature03597}, \href
  {https://ui.adsabs.harvard.edu/abs/2005Natur.435..629S} {435, 629}

\bibitem[\protect\citeauthoryear{{Tang} et~al.,}{{Tang} et~al.}{2019}]{Tang19}
{Tang} J.-J.,  et~al., 2019, \mn@doi [\mnras] {10.1093/mnras/stz134}, \href
  {https://ui.adsabs.harvard.edu/abs/2019MNRAS.484.2575T} {484, 2575}

\bibitem[\protect\citeauthoryear{{Trapp} et~al.,}{{Trapp}
  et~al.}{2022}]{Trapp2021}
{Trapp} C.~W.,  et~al., 2022, \mn@doi [\mnras] {10.1093/mnras/stab3251}, \href
  {https://ui.adsabs.harvard.edu/abs/2022MNRAS.509.4149T} {509, 4149}

\bibitem[\protect\citeauthoryear{{Tremaine} et~al.,}{{Tremaine}
  et~al.}{2002}]{Tremaine02}
{Tremaine} S.,  et~al., 2002, \mn@doi [\apj] {10.1086/341002}, \href
  {https://ui.adsabs.harvard.edu/abs/2002ApJ...574..740T} {574, 740}

\bibitem[\protect\citeauthoryear{{Ueda}, {Akiyama}, {Ohta}  \& {Miyaji}}{{Ueda}
  et~al.}{2003}]{Ueda03}
{Ueda} Y.,  {Akiyama} M.,  {Ohta} K.,   {Miyaji} T.,  2003, \mn@doi [\apj]
  {10.1086/378940}, \href
  {https://ui.adsabs.harvard.edu/abs/2003ApJ...598..886U} {598, 886}

\bibitem[\protect\citeauthoryear{Veilleux et~al.,}{Veilleux
  et~al.}{2023}]{Veilleux2023}
Veilleux S.,  et~al., 2023, \mn@doi [The Astrophysical Journal]
  {10.3847/1538-4357/ace10f}, 953, 56

\bibitem[\protect\citeauthoryear{{Vogelsberger}, {Marinacci}, {Torrey}  \&
  {Puchwein}}{{Vogelsberger} et~al.}{2020}]{Vogelsberger2020}
{Vogelsberger} M.,  {Marinacci} F.,  {Torrey} P.,   {Puchwein} E.,  2020,
  \mn@doi [Nature Reviews Physics] {10.1038/s42254-019-0127-2}, \href
  {https://ui.adsabs.harvard.edu/abs/2020NatRP...2...42V} {2, 42}

\bibitem[\protect\citeauthoryear{{Volonteri}, {Capelo}, {Netzer}, {Bellovary},
  {Dotti}  \& {Governato}}{{Volonteri} et~al.}{2015}]{Volonteri2015b}
{Volonteri} M.,  {Capelo} P.~R.,  {Netzer} H.,  {Bellovary} J.,  {Dotti} M.,
  {Governato} F.,  2015, \mn@doi [\mnras] {10.1093/mnras/stv387}, \href
  {https://ui.adsabs.harvard.edu/abs/2015MNRAS.449.1470V} {449, 1470}

\bibitem[\protect\citeauthoryear{{Volonteri}, {Habouzit}  \&
  {Colpi}}{{Volonteri} et~al.}{2021}]{Volonteri21}
{Volonteri} M.,  {Habouzit} M.,   {Colpi} M.,  2021, \mn@doi [Nature Reviews
  Physics] {10.1038/s42254-021-00364-9}, \href
  {https://ui.adsabs.harvard.edu/abs/2021NatRP...3..732V} {3, 732}

\bibitem[\protect\citeauthoryear{{Wang} et~al.,}{{Wang} et~al.}{2021}]{Wang21}
{Wang} F.,  et~al., 2021, \mn@doi [\apjl] {10.3847/2041-8213/abd8c6}, \href
  {https://ui.adsabs.harvard.edu/abs/2021ApJ...907L...1W} {907, L1}

\bibitem[\protect\citeauthoryear{Wechsler \& Tinker}{Wechsler \&
  Tinker}{2018}]{Wechsler2018}
Wechsler R.~H.,  Tinker J.~L.,  2018, \mn@doi [Annual Review of Astronomy and
  Astrophysics] {10.1146/annurev-astro-081817-051756}, 56, 435

\bibitem[\protect\citeauthoryear{{Weinberger} et~al.,}{{Weinberger}
  et~al.}{2018}]{Weinberger18}
{Weinberger} R.,  et~al., 2018, \mn@doi [\mnras] {10.1093/mnras/sty1733}, \href
  {https://ui.adsabs.harvard.edu/abs/2018MNRAS.479.4056W} {479, 4056}

\bibitem[\protect\citeauthoryear{{Yang} et~al.,}{{Yang} et~al.}{2020}]{Yang20}
{Yang} J.,  et~al., 2020, \mn@doi [\apjl] {10.3847/2041-8213/ab9c26}, \href
  {https://ui.adsabs.harvard.edu/abs/2020ApJ...897L..14Y} {897, L14}

\bibitem[\protect\citeauthoryear{{Yu} \& {Tremaine}}{{Yu} \&
  {Tremaine}}{2002}]{Yu02}
{Yu} Q.,  {Tremaine} S.,  2002, \mn@doi [\mnras]
  {10.1046/j.1365-8711.2002.05532.x}, \href
  {https://ui.adsabs.harvard.edu/abs/2002MNRAS.335..965Y} {335, 965}

\bibitem[\protect\citeauthoryear{{Yuan} \& {Narayan}}{{Yuan} \&
  {Narayan}}{2014}]{Yuan14}
{Yuan} F.,  {Narayan} R.,  2014, \mn@doi [\araa]
  {10.1146/annurev-astro-082812-141003}, \href
  {https://ui.adsabs.harvard.edu/abs/2014ARA&A..52..529Y} {52, 529}

\makeatother
\end{thebibliography}




\appendix
\section{Derivation of transition $\alpha$ that separates two phases of accretion}
\label{sec_app}

The general solution for the flow onto a central black hole in a Hernquist halo (Eq.~\ref{eq_hern_bh_sol}) can be written as, 
\begin{equation}
   \frac{\mathcal{M}^2}{2} - \ln{\mathcal{M} x^2} - \frac{1}{x + \alpha} - \frac{\beta}{x} = C
   \label{eq_ap_sol}
\end{equation}
where $x=r/l_{\rm h}$, $\alpha=a/l_{\rm h}$ and $\beta=M_{\rm bh}/M_{\rm h}$.
For $\alpha>\alpha_{\rm T}$, the trans-sonic solution passes through only the innermost critical point, and for $\alpha<\alpha_{\rm T}$, the trans-sonic solution passes through only the outermost critical point. At the transition, $\alpha=\alpha_{\rm T}$, the solution passes through both the critical points.  

For the trans-sonic solution passing through the outermost critical point, $\mathcal{M}=1$, $x \approx 0.5$, and as $\alpha\ll 0.5$ and $\beta \ll 1$,  Eq.~\ref{eq_ap_sol} yields, $C \approx C_{\rm T}$, where,
\begin{equation}
    C_{\rm T} = 2\ln{2} - 3/2 \ .
\end{equation}
At $\alpha=\alpha_{\rm T}$,  the same trans-sonic solution also passes through the innermost critical point, for which we can substitute $\mathcal{M}=1$, $x \approx \beta/2$ and $C=C_{\rm T}$ in Eq.~\ref{eq_ap_sol} to obtain,
\begin{equation}
   \frac{1}{2} - 2\ln{\beta} + \ln{4} - \frac{2}{\beta+2\alpha} - 2 =  2\ln{2} - 3/2
\end{equation}
which for $\beta\ll \alpha$ gives, 
\begin{equation}
    \alpha_{\rm T} \approx \frac{1}{2\ln(1/\beta)} \ ,
    \label{eq_transition_trasnsition}
\end{equation}
that is Eq.~\ref{eq_turnover}. We note that $\beta$ is the ratio of the black hole mass and the halo mass and $\beta\ll 1$. Repeating the same procedure for the NFW halo gives (as in Eq.~\ref{eq_turnover_nfw}),
\begin{equation}
\alpha_{\rm T} \approx \frac{1}{2 \eta \ln(1/\beta)} \ ,
\end{equation}
where $\eta = (1+1/c)\ln(1+c)-1$.  For $c=5$, $\eta \approx 1$;  and for $c=15$, $\eta\approx 2$. 

The expressions for $\alpha_{\rm T}$ can also be derived by equating the mass accretion rates in two asymptotic regimes; by equating the mass accretion rates in equations~\ref{eq_mass_acc_hern} and \ref{eq_mass_acc_hern2} for the Hernquist halo, and by equating the mass accretion rates in Eq.~\ref{eq_lamb_nfw} for the NFW halo.




\bsp	
\label{lastpage}
\end{document}